\documentclass[%
 reprint,
superscriptaddress,
 amsmath,amssymb,
 aps,
]{revtex4-2}

\usepackage{graphicx}
\usepackage{dcolumn}
\usepackage{bm}
\usepackage{float}
\usepackage{mathtools}
\usepackage{braket}
\usepackage{amsmath}
\usepackage{amstext}
\usepackage{xcolor}
\usepackage{caption}
\usepackage{array}
\usepackage{gensymb}
\usepackage{appendix}

\begin{document}

\preprint{APS/123-QED}

\title{A Unified SU(2) Framework for Vector Beam Transformations \\ and Complex Beam Shaping}

\author{Gayathri G T}
 
 \affiliation{Materials Science Group, Indira Gandhi Centre for Atomic Research, Kalpakkam- 603102, India}
 \affiliation{Homi Bhabha National Institute, Training School Complex, Anushakti Nagar, Mumbai- 400094, India}

\author{Gururaj Kadiri}
 \email{gururaj@igcar.gov.in}
 \affiliation{Materials Science Group, Indira Gandhi Centre for Atomic Research, Kalpakkam- 603102, India}
 \affiliation{Homi Bhabha National Institute, Training School Complex, Anushakti Nagar, Mumbai- 400094, India}

\date{\today}

\begin{abstract}
We present a constructive framework for designing transformations between structured light fields using birefringent optical elements, formulated in terms of SU(2) operations on polarization. Within this framework, transformations between vector beams are treated as spatially varying SU(2) operations, leading to a direct procedure for designing doubly inhomogeneous waveplates (d-plates) that implement the desired mapping. We identify a condition under which a single element implements a prescribed transformation exactly, including the global phase, and provide an explicit prescription for constructing the corresponding doubly inhomogeneous waveplate (d-plate) when this condition is satisfied, along with its realization using three singly inhomogeneous waveplates. Within this formulation, a broad class of problems in structured light can be treated within a single framework, including vector beam transformations, spin-orbital dynamics, and complex beam shaping. Crucially, the same SU(2) operations directly realize quantum channels on the orbital angular momentum degree of freedom, with polarization serving as a physical ancilla. These results establish a unified and explicitly constructive route to complex beam shaping and vector beam transformations based on SU(2) parameter synthesis, and provide a systematic foundation for designing next-generation photonic elements for structured light and spin-orbit information processing.
\end{abstract}

\maketitle


\section{\label{Sec:Introduction}Introduction}

Structured light beam refers to light beam with spatial inhomogeneity in its phase, amplitude and polarization across the transverse plane \cite{andrews2011structured,he2022towards,rubinsztein2016roadmap,angelsky2020structured}. The process of generating such structured light beams from a standard light beam is referred to as complex beam shaping \cite{dickey2018laser}. It finds applications in various domains that include optical trapping \cite{woerdemann2013advanced}, optical communication \cite{al2021structured}, laser welding \cite{chen2026review} and optical imaging for biomedical applications \cite{angelo2019review}. Vector beams are also structured light beams with spatially varying states of polarization (SoP) across their transverse plane \cite{zhan2009cylindrical}. In case of higher dimensions, a beam that is inhomogeneous in its Orbital Angular Momentum (OAM) degree of freedom is also a vector beam. OAM plays a major role in the implementation of certain techniques or algorithms such as quantum walks \cite{venegas2012quantum,portugal2013quantum}. These vector beams are generated using liquid-crystal devices like spatial light modulators \cite{neil2002method,maurer2007tailoring,chen2015complete} and q-plates \cite{marrucci2006optical,piccirillo2010photon,d2015arbitrary,vergara2018generalized}, holograms \cite{fu2017tailoring,ruiz2013highly} and interferometric techniques \cite{hall1996vector,tidwell1990generating,passilly2005simple}. In recent years, metasurfaces \cite{arbabi2015dielectric,wang2021efficient,wu2022dielectric,liu2022generation,chen2016review} have emerged as most favoured candidate for complex beam shaping tasks. A subclass of metasurfaces called birefringent metasurfaces \cite{balthasar2017metasurface,mirzapourbeinekalaye2022general} are used with respect to vector beam generation. These are Pancharatnam Berry Phase Optical elements (PBOEs) whose action on an incident light beam induces a phase difference between the two orthogonal plane-polarized components. These plane-polarized components between which the phase is introduced is, in general, spatially-varying. Here we extend the definition of birefringent metasurfaces so as to allow even the magnitude of imparted phase to vary spatially. Such generalized birefringent metasurfaces exhibit the same action as another class of PBOEs called the doubly-inhomogeneous waveplates (d-plates) \cite{radhakrishna2021realization}. Waveplates are optical elements characterized by two parameters: retardance and fast-axis orientation. If both the parameters are varying spatially, then the waveplates are referred to as d-plates. These d-plates constitute a powerful platform for structured light engineering. Whereas, if a waveplate has either of the two parameters spatially varying, it is referred to as singly-inhomogeneous or s-plates. Unlike these singly inhomogeneous elements such as the q-plate, a d-plate enables concurrent control over dynamic phase and geometric (Pancharatnam-Berry) phase, thereby allowing full manipulation of polarization and phase degrees of freedom. Such devices facilitate the generation of vector vortex beams and full Poincar\'e beams with spatially varying states of polarization, as well as controlled spin-orbit angular momentum conversion. This enhanced degree of freedom is advantageous in various applications as mentioned above, spanning polarization-sensitive imaging, optical trapping, quantum state engineering in hybrid spin-orbital Hilbert spaces, and optical communication. By coupling spatial phase, polarization topology, and angular momentum in a single compact element, d-plates provide a unified framework for tailoring the complete electromagnetic field structure, positioning them as key components in advanced photonic and quantum optical systems. \citeauthor{radhakrishna2021realization} in \cite{radhakrishna2021realization} have realized the equivalence of d-plate involving only s-plates by using a QHQ-s-plate gadget. This has been done by performing local manipulation of the amplitude and phase of light beams. They have illustrated this idea towards two different applications. One is by tailoring the complex amplitude of light beams to simulate Laguerre-Gaussian (LG) and Hermite-Gaussian (HG) beams of higher order and the other is by designing a d-plate such that it is a polarization-dependent spatially varying phase plate. This paper focuses mainly on the study of vector beam transformation using d-plates and the application of d-plates in complex beam shaping.
\par
The paper is organized as follows: in Section \ref{Sec:Theory}, we give a theoretical background on the transformation between vector beams associated with polarization and OAM states of light using waveplates and spiral phase plates (SPPs). Here, we also give a brief description of quantum walks and its implementation in the optical regime. In Section \ref{Sec:VBT}, we have studied vector beam transformations using single and multiple waveplates and its equivalence to d-plate. In Section \ref{Sec:CBS} we explore the use of d-plates in complex beam shaping tasks. Section \ref{Sec:Illustrations} gives some illustrations on the generation and transformation of vector beams using d-plates which finds applications in quantum walks, transformation between full Poincar\'e beams and complex beam shaping. We conclude the paper by summarizing the results in Section \ref{Sec:Conclusion}. Illustrations of the action of d-plate and waveplate on vector beams is depicted in Appendix.

\section{\label{Sec:Theory}Theoretical background}

An arbitrary state of polarization (SoP) of light beam can be written as,
\begin{equation}
    \ket{\boldsymbol{s}} = s_h \ket{\boldsymbol{h}} + s_v \ket{\boldsymbol{v}},
\label{ScalarBeam}
\end{equation}
where $\{\ket{\boldsymbol{h}}, \ket{\boldsymbol{v}}\}$ constitute a 2-dimensional basis with $\ket{\boldsymbol{h}}$ and $\ket{\boldsymbol{v}}$ representing the horizontal and vertical SoPs. $s_h$ and $s_v$ are complex numbers such that $|s_h|^2 + |s_v|^2 = 1$. The SoP defined in Eq.~\eqref{ScalarBeam} does not vary spatially and are called scalar beams. The state orthogonal to the SoP $\ket{\boldsymbol{s}}$ is defined as,
\begin{equation}
    \ket{\boldsymbol{s}_\perp} = -\bar{s_v} \ket{\boldsymbol{h}} + \bar{s_h} \ket{\boldsymbol{v}}
\end{equation}
where $\bar{s_h}$ and $\bar{s_v}$ denotes the complex conjugate of $s_h$ and $s_v$ respectively.
\par
Light beams can have spatially varying SoPs, known as vector beams. The state of an arbitrary vector beam can be defined similar to Eq.~\eqref{ScalarBeam} as,
\begin{equation}
    \ket{\boldsymbol{s}(\vec{\boldsymbol{r}})} = s_h(\vec{\boldsymbol{r}}) \ket{\boldsymbol{h}} + s_v (\vec{\boldsymbol{r}})\ket{\boldsymbol{v}},
\label{VectorBeam}
\end{equation}
where the coefficients $s_h(\vec{\boldsymbol{r}})$ and $s_v(\vec{\boldsymbol{r}})$ are functions of the spatial vector $\vec{\boldsymbol{r}} = (r,\phi)$, that is, it varies along the radial direction $r$ and azimuthal direction $\phi$, satisfying the relation $|s_h(\vec{\boldsymbol{r}})|^2 + |s_v(\vec{\boldsymbol{r}})|^2 = 1$. Another case of a vector beam is, light beams comprising of multiple OAM components. We represent the vector beams having a definite OAM by the symbol $\ket{\boldsymbol{s};m}$, and they are defined as, 
\begin{equation}
    \ket{\boldsymbol{s};m} = \ket{\boldsymbol{s}}\otimes\ket{m}_o = \ket{\boldsymbol{s}}e^{im\phi}
\label{OAM}
\end{equation}
where $\ket{\boldsymbol{s}}$ represents the SoP and the states $\{\ket{m}_o, m=..-1,0,1,..\}$ constitute the eigenstates of the OAM operator. These states show up as azimuthal phase dependence $e^{im\phi}$. Now, light beams consisting of several OAM components, termed as vector beams can be expressed as: 
\begin{equation}
    \ket{\boldsymbol{S}} = \sum_{m=b}^e s_m \ket{\boldsymbol{s}_m; m}
\label{OAMComposite}
\end{equation}
where $s_m$ are complex numbers such that $\sum_m |s_m|^2 = 1$, and $b$ and $e\geq b$ are integers, indicating the span of the OAM terms involved in the superposition. For $\ket{\boldsymbol{S}}$ to represent a vector beam, atleast two of the SoPs $\ket{\boldsymbol{s}_m}$ must be distinct. 
\par
Transformations between SoPs without loss of intensity are carried out using SU(2) operators denoted by $\hat{S}_\Gamma(\boldsymbol{k})$. These operators are characterized by two parameters: a scalar $\Gamma$ and a unit vector $\boldsymbol{k}$ in 3D: 
\begin{equation}
    \hat{S}_\Gamma(\boldsymbol{k}) = e^{-i\frac{\Gamma}{2}(\boldsymbol{k}\cdot\boldsymbol{\sigma})} = \cos\frac{\Gamma}{2} \hat{I} -i\sin\frac{\Gamma}{2} (\boldsymbol{k} \cdot \boldsymbol{\sigma})
\label{SU2}
\end{equation}
where $\boldsymbol{k}=(k_x, k_y, k_z)$ is a unit vector in 3D, $\hat{I}$ is the identity operator and $\boldsymbol{\sigma}=(\hat{\sigma}_x, \hat{\sigma}_y, \hat{\sigma}_z)$ is a vector of $2\times2$ matrices called Pauli matrices, defined as,
\begin{equation}
\begin{aligned}
    \hat{\sigma}_x &= \ket{\boldsymbol{h}}\bra{\boldsymbol{h}}-\ket{\boldsymbol{v}}\bra{\boldsymbol{v}}, \\ 
    \hat{\sigma}_y &= \ket{\boldsymbol{d}}\bra{\boldsymbol{d}}-\ket{\boldsymbol{a}}\bra{\boldsymbol{a}}, \ \text{and} \\ 
    \hat{\sigma}_z &= \ket{\boldsymbol{l}}\bra{\boldsymbol{l}}-\ket{\boldsymbol{r}}\bra{\boldsymbol{r}}.
\end{aligned}
\label{PauliMatrix}
\end{equation}
The scalar SoPs $\ket{\boldsymbol{d}}$, $\ket{\boldsymbol{a}}$, $\ket{\boldsymbol{l}}$, $\ket{\boldsymbol{r}}$ stand for diagonal, anti-diagonal, left circular and right circular SoPs respectively, and are defined as, $\ket{\boldsymbol{d}} \equiv \frac{1}{\sqrt{2}}(\ket{\boldsymbol{h}}+\ket{\boldsymbol{v}})$, $\ket{\boldsymbol{a}} \equiv \frac{1}{\sqrt{2}}(\ket{\boldsymbol{v}}-\ket{\boldsymbol{h}})$, $\ket{\boldsymbol{l}} \equiv \frac{1}{\sqrt{2}}(\ket{\boldsymbol{h}}+i \ket{\boldsymbol{v}})$ and $\ket{\boldsymbol{r}} \equiv \frac{1}{\sqrt{2}}(i \ket{\boldsymbol{h}}+\ket{\boldsymbol{v}})$. 
The definition of the Pauli operators in Eq.~\eqref{PauliMatrix} warrants particular attention. In the standard treatment, the Pauli matrices are introduced as explicit $2\times 2$ matrices in a fixed basis, with the 
physical identification with polarization states made subsequently through a specific coordinate representation. Here, by contrast, the Pauli operators are defined directly through their spectral decomposition in terms of physically meaningful polarization states, without invoking any matrix representation. Each operator is expressed as the difference of projectors onto a pair of orthogonal polarization states, where the three pairs $\{\ket{\boldsymbol{h}},\ket{\boldsymbol{v}}\}$, $\{\ket{\boldsymbol{d}}, \ket{\boldsymbol{a}}\}$ and $\{\ket{\boldsymbol{l}}, \ket{\boldsymbol{r}}\}$ correspond to the three orthogonal axes of the Poincar\'{e} sphere. This definition is basis-independent: the operators $\hat{\sigma}_x$, $\hat{\sigma}_y$, $\hat{\sigma}_z$ are physical objects whose identity is determined entirely by the geometry of the Poincar\'{e} sphere, independent of any choice of coordinate representation. 
Note that in the standard $\{|h\rangle, |v\rangle\}$ matrix representation, these operators correspond to a cyclic permutation of the conventional Pauli matrices $(\hat{\sigma}_x, \hat{\sigma}_y, \hat{\sigma}_z) \equiv (\sigma_z, \sigma_x, \sigma_y)$, which leaves the commutation relations $[\hat{\sigma}_i, \hat{\sigma}_j] = 2i\varepsilon_{ijk}\hat{\sigma}_k$ and all SU(2) properties invariant.
However, as will be seen in Section~\ref{Sec:VBT}, this basis-independent definition of the operators has direct consequences for the constructive design procedure for optical elements developed therein. 
\par
In the SU(2) transformation $\hat{S}_\Gamma(\boldsymbol{k)}$, we refer to the $\Gamma$ as the "rotation angle" and $\boldsymbol{k}$ as the "rotation axis". This nomenclature stems from the fact that an SU(2) matrix acts as a rotation operator on the Stokes vector representation of the SoP, which for an SoP $\ket{\boldsymbol{u}}$ is defined as:
\begin{equation}
    \hat{s}(\ket{\boldsymbol{u}})_i=\braket{\boldsymbol{u}|\hat{\sigma}_i|\boldsymbol{u}}, i=x,y,z
    \label{StokesVector}
\end{equation}
where $\hat{\sigma}_i$ are the operators defined in Eq.~\eqref{PauliMatrix}. Alternatively, the action of an SU(2) operator on a state of polarization can be understood as imparting a definite phase difference between two of its orthogonal components. For transformation between SoPs of vector beams, the $\Gamma$ and $\boldsymbol{k}$ of the SU(2) operator varies spatially with respect to $\vec{\boldsymbol{r}}$ as, $\hat{S}_{\Gamma(\vec{\boldsymbol{r}})}(\boldsymbol{k}(\vec{\boldsymbol{r}}))$. 
\par
Physically, the transformations between SoPs are realized in optical systems using birefringent materials called waveplates. Waveplates are SU(2) operators with their $\boldsymbol{k}$ confined to the $X-Y$ plane. So, $\boldsymbol{k}=(k_x, k_y, 0)$ with $k_x=\cos 2\alpha$ and $k_y=\sin 2\alpha$. A waveplate is represented by the symbol, $\hat{W}_\Gamma(\alpha)$ where $\alpha$ is half of the angle that $\boldsymbol{k}$ makes with the x-axis. Here, $\Gamma$ is termed as its retardance and $\alpha$ as the orientation of its fast-axis \cite{kumar2011polarization}. The action of a waveplate is defined from Eq.~\eqref{SU2} as,
\begin{equation}
    \hat{W}_\Gamma(\alpha) = \cos\frac{\Gamma}{2} \hat{I} -i\sin\frac{\Gamma}{2} (\cos 2\alpha \cdot \hat{\sigma}_x + \sin 2\alpha \cdot \hat{\sigma}_y).
\end{equation}
With respect to waveplates, the circular SoPs $\ket{\boldsymbol{l}}$ and $\ket{\boldsymbol{r}}$ stand on a different footing compared to the rest of the SoPs, in the sense that every fast-axis of the waveplate can transform $\ket{\boldsymbol{l}}$ to $\ket{\boldsymbol{r}}$ and vice-versa, up to a phase. No other pair of orthogonal SoPs display this behavior. 

Like with SU(2) transformations, the waveplates can also have retardance and fast-axis orientation varying spatially with $\vec{\boldsymbol{r}}$. We shall refer to the waveplates in which both the retardance $\Gamma$ and fast-axis orientation $\alpha$ vary spatially as the "doubly-inhomogeneous" waveplates or d-plates and "singly-inhomogeneous" waveplates or s-plates, if only one of them varies spatially. A well-known class of such singly inhomogeneous waveplates is that of q-plates \cite{kadiri2019wavelength,rubano2019q,hakobyan2025q} . A q-plate is defined in terms of its action on the circular SoPs as,
\begin{equation}
    \begin{aligned}
        \hat{q}_\delta(m) \ket{\boldsymbol{l};g} &= e^{i\delta}\ket{\boldsymbol{r};g+m}, \ \text{and} \\
         \hat{q}_\delta(m) \ket{\boldsymbol{r};g} &= -e^{-i\delta}\ket{\boldsymbol{l};g-m}
    \end{aligned}
\label{qplateAction}
\end{equation}
Here, $m$ is an integer indicating the amount of OAM induced by the plate and $\delta$ is a relative phase factor that controls the phase difference between the left and right circular SoPs at the zero azimuth. The action of the q-plate $\hat{q}_\delta(m)$ on a state with any other SoP $\ket{\boldsymbol{s};g}$ can be obtained by expressing $\ket{\boldsymbol{s}}$ as a linear combination of $\ket{\boldsymbol{l}}$ and $\ket{\boldsymbol{r}}$, and acting $\hat{q}_\delta(m)$ linearly on them. The emerging beam in this case is a vector beam having an azimuthaly varying SoP. The waveplate $\hat{W}_\Gamma(\alpha)$ that accomplishes the above transformation in Eq.~\eqref{qplateAction} has $\Gamma=\pi$ and $\alpha$ varying linearly with the azimuthal angle $\phi$ as $\alpha=q\phi + \alpha_0$ where $q$, $\alpha_0$ are constants and $q$ is related to the magnitude of OAM obtained as $m=2q$. These q-plates have been playing a central role in implementing quantum walks on the OAM states of light beams. 
\par
Quantum walks are quantum mechanical analogue of classical random walks \cite{venegas2012quantum,portugal2013quantum}. They have been employed in understanding many physical phenomena \cite{kadian2021quantum,tanaka2022spatial}. It has been implemented in various physical platforms such as NMR quantum information processor \cite{ryan2005experimental}, trapped ions \cite{zahringer2010realization} and single photons \cite{broome2010discrete}. Here we shall restrict our attention to coin-based discrete-time quantum walks (DTQWs). A single step of such a quantum walk comprises of two operations and is defined as: 
\begin{equation}
    \hat{T} = \hat{S}(\hat{C} \otimes \hat{I}).
\end{equation}
where $\hat{C}$ is the coin toss operator that places the coin in a superposition of heads and tails and $\hat{S}$ is the shift operator that moves the position of the walker forward or backward depending on the output of the coin toss. For a standard quantum step, the shift operator shifts the position of the walker by $\pm1$ unit as given by the following equation.
\begin{equation}
    \begin{aligned}
        \hat{T}(\boldsymbol{c},\boldsymbol{s})\ket{\boldsymbol{s};g} &= \ket{\boldsymbol{c};g+1} \\
        \hat{T}(\boldsymbol{c},\boldsymbol{s})\ket{\boldsymbol{s}_\perp;g} &= \ket{\boldsymbol{c}_\perp;g-1}
    \end{aligned}
\label{stdquantumstep}
\end{equation}
A quantum walk $\hat{W}$ of $N$ steps is of the form,
\begin{equation}
    \hat{W} = \hat{T}_N \dots \hat{T}_1
\end{equation}
The action of an arbitrary quantum walk $\hat{W}$ on the orthogonal pair of states $\ket{\boldsymbol{w};0}$ and $\ket{\boldsymbol{w}_\perp;0}$ is given by,
\begin{equation}
    \begin{aligned}
        \hat{W} \ket{\boldsymbol{w};0} &= \ket{\boldsymbol{W}} \\
        \hat{W} \ket{\boldsymbol{w}_\perp;0} &= \ket{\boldsymbol{W}_\perp}
    \end{aligned}
\end{equation}
where the states $\ket{\boldsymbol{W}}$ and $\ket{\boldsymbol{W}_\perp}$ are orthogonal to each other and are of the form Eq.~\eqref{OAMComposite}. 
\par
In optical implementation, the polarization degree of freedom serves as the coin space, while the OAM degree of freedom serves as the walk space \cite{zhang2010implementation,goyal2012implementing,neves2018photonic}. Coin toss operator corresponds to placing the circular SoPs into their superpositions \cite{chandrashekar2008optimizing}, and is accomplished through the use of single or multiple waveplates. Shift operator is implemented using inhomogeneous waveplates called the q-plates which moves the walker forward in the OAM space if the SoP is left-circular, and backward if it is right-circular as given in Eq.~\eqref{qplateAction}. 
\par
A single step of a quantum walk is implemented by employing a waveplate followed by a q-plate:
\begin{equation}
    \hat{T}_\delta(m, \Gamma, \alpha) = \hat{q}_\delta(m)\hat{W}_\Gamma(\alpha)
\end{equation}
The waveplate $\hat{W}_\Gamma(\alpha)$ functions like the coin toss operator, placing the SoPs $\ket{\boldsymbol{l}}$ and $\ket{\boldsymbol{r}}$ into their superpositions. The q-plate $\hat{q}_\delta(m)$ advances the OAM of these components by $\pm m$ units, in addition to flipping the polarizations. 
\par
In addition to q-plate the OAM states of the form Eq. \eqref{OAM} can also be generated using Spiral Phase Plate. These plates are made of transparent material, the thickness of which varies with the azimuthal angle, thereby imparting a helical phase \cite{beijersbergen1994helical}. The action of the SPP represented as $\hat{Q}_m$ on the scalar SoP $\ket{\boldsymbol{s}}$ of the form Eq.~\eqref{ScalarBeam} is given as: 
\begin{equation}
    \hat{Q}_m\ket{\boldsymbol{s}} = e^{im\phi}\ket{\boldsymbol{s}} \equiv \ket{\boldsymbol{s}; m}
\label{SPP}
\end{equation}
where $m$ is the topological charge of the SPP indicating the magnitude of OAM and $\phi$ represents the azimuthal angle coordinate. So, the SPP introduces a helical phase front to the beam, which corresponds to imparting an OAM of magnitude $m$.
\par
For an arbitrary transformation between the SoPs or OAM states, it is essential to determine the parameters $\Gamma$ and $\boldsymbol{k}$ of the SU(2) transformation and therefore the parameters of the waveplate. In the following sections, we determine the parameters of the waveplate for vector beam transformations and discuss the application of d-plate in complex beam shaping.

\section{\label{Sec:VBT}Vector beam transformations}

Consider two scalar SoPs $\ket{\boldsymbol{u}}$ and $\ket{\boldsymbol{w}}$ of the form Eq.~\eqref{ScalarBeam}. The central result of this paper is that it is always possible to identify an SU(2) transformation $\hat{S}_\Gamma(\boldsymbol{k})$ of the form Eq.~\eqref{SU2} that transforms the former to the latter:
\begin{equation}
    \ket{\boldsymbol{w}} = \hat{S}_\Gamma(\boldsymbol{k}) \ket{\boldsymbol{u}}
\end{equation}
The parameters $\Gamma$ and $\boldsymbol{k}$ of the SU(2) transformation, depends on the real and imaginary terms of the four complex inner products $\braket{\boldsymbol{u}|\boldsymbol{w}}$ and $\braket{\boldsymbol{u}|\hat{\sigma}_i|\boldsymbol{w}}$, for $i=x, y$ and $z$:
\begin{equation}
    \begin{aligned}
        \cos\frac{\Gamma}{2} &= \textbf{re}(\braket{\boldsymbol{u}|\boldsymbol{w}}), \\
        k_i \sin\frac{\Gamma}{2} &= -\textbf{imag}(\braket{\boldsymbol{u}|\hat{\sigma}_i|\boldsymbol{w}}) \ \text{for} \ i=x, y, z,
    \end{aligned}
\label{SU2Parameters}
\end{equation}
where $\braket{\boldsymbol{u}|\boldsymbol{w}}$ refers to the inner product of the two SoPs, $\ket{\boldsymbol{u}}$ and $\ket{\boldsymbol{w}}$, \textbf{re}(c) and \textbf{imag}(c) stand for the real and imaginary parts of the complex number c. These SU(2) transformations can be realized using waveplates as mentioned in Section \ref{Sec:Theory}.

\subsection{\label{Sec:waveplate}Using single waveplate}

We now discuss the case of designing a waveplate for transforming a given pair of beams. Given two scalar SoPs $\ket{\boldsymbol{u}}$ and $\ket{\boldsymbol{w}}$, there always exists a waveplate $\hat{W}_\Gamma(\alpha)$ that transforms $\ket{\boldsymbol{u}}$ to $\ket{\boldsymbol{w}}$, up to a phase. The retardance $\Gamma$ and fast axis orientation $\alpha$ of such a waveplate follow from Eq.~\eqref{SU2Parameters} and are given as: 
\begin{equation}
    \begin{aligned}
        \Gamma &= 2\cos^{-1}\ (r_0 \cos\ (\phi_0 - \phi_z)), \\
        \alpha &= \frac{1}{2} \tan^{-1} \Bigg(\frac{r_y\sin(\phi_y-\phi_z)}{r_x\sin(\phi_x-\phi_z)}\Bigg)
    \end{aligned}
\label{SingleWP_dplate}
\end{equation}
where $r_0, r_i, \phi_0$ and $\phi_i$ are defined as:
\begin{equation}
    \begin{aligned}
        r_0 &= |\braket{\boldsymbol{u}|\boldsymbol{w}}|; & r_i&=|\braket{\boldsymbol{u}|\hat{\sigma}_i|\boldsymbol{w}}| & \text{for} \ i=x, y; \\
        \phi_0&=\Phi(\braket{\boldsymbol{u}|\boldsymbol{w}}); & \text{and}\  \phi_i&=\Phi(\braket{\boldsymbol{u}|\hat{\sigma}_i|\boldsymbol{w}}) & \text{for} \ i=x, y, z
    \end{aligned}
\end{equation}
where $\Phi(c)$ denotes the phase of the complex number $c$. This waveplate takes $\ket{\boldsymbol{u}}$ to $e^{-i\phi_z}\ket{\boldsymbol{w}}$ where $\phi_z(\boldsymbol{u},\boldsymbol{w})=\Phi(\braket{\boldsymbol{u}|\hat{\sigma}_z|\boldsymbol{w}})$, $\sigma_z$ is the Pauli matrix defined in Eq. ~\eqref{PauliMatrix}. It is therefore possible to transform the SoP $\ket{\boldsymbol{u}}$ to the SoP $\ket{\boldsymbol{w}}$ exactly without any additional phase factors, provided $\phi_z(\boldsymbol{u},\boldsymbol{w})\equiv\Phi(\braket{\boldsymbol{u}|\hat{\sigma}_z|\boldsymbol{w}})=2n\pi$ for some integer $n$. 
\par
The above result is true even for the case of vector beams. If the involved states are vector beams $\ket{\boldsymbol{u}(\vec{\boldsymbol{r}})}$ and $\ket{\boldsymbol{w}(\vec{\boldsymbol{r}})}$, the above constraint $\phi_z(\boldsymbol{u}(\vec{\boldsymbol{r}}),\boldsymbol{w}(\vec{\boldsymbol{r}}))=2n\pi$ must hold at every $\vec{\boldsymbol{r}}$: 
 \begin{equation}
     \begin{aligned}
         \text{if} \ \hat{W}_{\Gamma(\vec{\boldsymbol{r}})}(\alpha(\vec{\boldsymbol{r}}))\ \ket{\boldsymbol{u}(\vec{\boldsymbol{r}})} &= \ket{\boldsymbol{w}(\vec{\boldsymbol{r}})}, \ \text{then} \\
         \Phi(\braket{\boldsymbol{u}(\vec{\boldsymbol{r}})|\boldsymbol{l}} \braket{\boldsymbol{l}|\boldsymbol{w}(\vec{\boldsymbol{r}})} - \braket{\boldsymbol{u}(\vec{\boldsymbol{r}})|\boldsymbol{r}} \braket{\boldsymbol{r}|\boldsymbol{w}(\vec{\boldsymbol{r}})}) &= 2n\pi.
     \end{aligned}
 \label{Phase_condition}
 \end{equation}
The waveplate $\hat{W}_{\Gamma(\vec{\boldsymbol{r}})}(\alpha(\vec{\boldsymbol{r}}))$ is termed as the d-plate in which the retardance $\Gamma(\vec{\boldsymbol{r}})$ and the fast-axis orientation $\alpha(\vec{\boldsymbol{r}})$ varies spatially across the waveplate. 
When the single-plate condition is not met, the transformation can still be realized exactly using two or three waveplates, as detailed in Section \ref{sec:Multiple_waveplates}.
\par
It is the basis-independent definition of the Pauli operators in Eq. ~\eqref{PauliMatrix} that endows the expressions in Eqs. ~\eqref{SU2Parameters} and~\eqref{SingleWP_dplate} with their coordinate-free character. The inner products $\braket{\boldsymbol{u}|\hat{\sigma}_i|\boldsymbol{w}}$ appearing in Eq.~\eqref{SU2Parameters} are basis-independent 
scalar quantities that encode the geometric relationship between the source and target states $\ket{\boldsymbol{u}}$ and $\ket{\boldsymbol{w}}$ on the Poincar\'e sphere, 
independent of any matrix representation. Consequently, 
Eqs. ~\eqref{SU2Parameters} and~\eqref{SingleWP_dplate} constitute a closed-form constructive solution for the waveplate parameters that depends only on the physical states $\ket{\boldsymbol{u}}$ and $\ket{\boldsymbol{w}}$, with no reference to any particular coordinate representation. This coordinate-free structure extends directly to the case of vector beams, where the parameters $\Gamma(\vec{\boldsymbol{r}})$ and $\alpha(\vec{\boldsymbol{r}})$ of the d-plate are obtained by 
evaluating Eq.~\eqref{SingleWP_dplate} pointwise, with the geometric content of the solution remaining transparent at every spatial point $\vec{\boldsymbol{r}}$. To our knowledge, this explicit basis-independent definition of the Pauli operators in terms of projectors onto the six cardinal polarization states, serving as the foundation for a closed-form constructive optical element design procedure, has not appeared previously in the polarization optics literature. 

\subsection{Using multiple waveplates}
\label{sec:Multiple_waveplates} 
Here, we claim that any SU(2) transformation $\hat{S}_\Gamma(\boldsymbol{k})$ can be realized by using multiple waveplates.  

\subsubsection{Two waveplates}

Any arbitrary SU(2) transformation can be realized using two waveplates $\hat{W}_{\Gamma_1}(\alpha_1)$ and $\hat{W}_{\Gamma_2}(\alpha_2)$ as:
\begin{equation}
   \hat{S}_\Gamma(\boldsymbol{k}) = \hat{W}_{\Gamma_2}(\alpha_2) \ \hat{W}_{\Gamma_1}(\alpha_1)
\label{SU2_twoWP}
\end{equation}
For a given $\hat{S}_\Gamma(\boldsymbol{k})$ there are infinite pairs of $\hat{W}_{\Gamma_1}(\alpha_1)$ and $\hat{W}_{\Gamma_2}(\alpha_2)$ that satisfy Eq. \eqref{SU2_twoWP}. So, the parameters of the two waveplates, if one of them is set at half-wave retardance is calculated to be:
\begin{equation}
    \begin{aligned}
        \Gamma_1 &= \pi, \\ 
        \cos\bigg({\frac{\Gamma_2}{2}}\bigg) &= \sqrt{k_x^2 + k_y^2} \ \sin\bigg({\frac{\Gamma}{2}}\bigg), \\
        \tan(2\alpha_1) &= \frac{k_y}{k_x}, \\
        \tan\big(2(\alpha_1 - \alpha_2)\big) &= -k_z \ \tan\bigg(\frac{\Gamma}{2}\bigg)
    \end{aligned}
\label{TwoWP_parameters}
\end{equation}

\subsubsection{Three waveplates}

In general, a set of three waveplates is sufficient to realize any arbitrary SU(2) transformation. In this case, the three waveplates $\hat{W}_{\Gamma_1}(\alpha_1)$, $\hat{W}_{\Gamma_2}(\alpha_2)$ and $\hat{W}_{\Gamma_3}(\alpha_3)$ need not be arbitrary. They can be set as two quarter-wave plates and one half-wave plate not arranged in any particular order \cite{simon1990minimal}. Here we study the combined action of these three waveplates, by placing them in the order $\hat{Q}_2\hat{Q}_1\hat{H}$. The SU(2) transformation realized using these three waveplates is given as:
\begin{equation}
   \hat{S}_\Gamma(\boldsymbol{k}) = \hat{W}_{\Gamma_2}(\alpha_2) \ \hat{W}_{\Gamma_1}(\alpha_1) \ \hat{W}_{\Gamma_H}(\alpha_H)
\end{equation}
So, the retardance of the three waveplates are fixed as $\Gamma_H=\pi$, $\Gamma_1=\frac{\pi}{2}$ and $\Gamma_2=\frac{\pi}{2}$. The parameters of the waveplates are then related as: 
\begin{equation}
    \begin{aligned}
        \sin(\alpha_2-\alpha_1) &= \sqrt{k_x^2+k_y^2} \ \sin\bigg(\frac{\Gamma}{2}\bigg) \\
        \tan(2\alpha_H + \alpha_2 - \alpha_1) &= -\frac{k_x}{k_y} \\
        \tan(\alpha_1 + \alpha_2 -2\alpha_H) &= k_z \ \tan\bigg(\frac{\Gamma}{2}\bigg)
    \end{aligned}
\label{ThreeWP_parameters}
\end{equation} 
where $\alpha_1$, $\alpha_2$ are the fast-axis orientations of the two quarter-wave plates $\hat{Q}_1$, $\hat{Q}_2$ and $\alpha_H$ is the fast-axis orientation of the half-wave plate $\hat{H}$. The above results in Eq.~\eqref{TwoWP_parameters} and \eqref{ThreeWP_parameters} hold even for the case of transformations between vector beams.

\section{\label{Sec:CBS} Complex beam shaping}

We have seen in Section \ref{Sec:waveplate} that a waveplate can have spatially varying retardance and fast-axis orientations. These non-standard waveplates called the "d-plates" are used to manipulate the phase and amplitude of the incident light beam, thus generating a vector beam. We now discuss the means of complex amplitude shaping using d-plates. \par
Consider a light beam with a uniform SoP $\ket{\boldsymbol{u}}$ and a complex-electric field distribution $E_i(\vec{\boldsymbol{r}})$. We seek to transform this light beam into one having same SoP $\ket{\boldsymbol{u}}$, but with a different complex-electric field distribution $E_f(\vec{\boldsymbol{r}})$. For this, however, we require the condition $E_i(\vec{\boldsymbol{r}}) \geq E_f(\vec{\boldsymbol{r}})$ to hold at every $\vec{\boldsymbol{r}}$ within the region of interest. Assuming it is so, we define a complex function $T(\vec{\boldsymbol{r}})$ as $T(\vec{\boldsymbol{r}}) = \frac{E_f(\vec{\boldsymbol{r}})}{E_i(\vec{\boldsymbol{r}})}$ such that $|T(\vec{\boldsymbol{r}})|\leq1$ at all $\vec{\boldsymbol{r}}$. With this function and the SoP $\ket{\boldsymbol{u}}$, we construct a vector beam $\ket{T(\vec{\boldsymbol{r}}), \boldsymbol{u}, \Delta}$ as:
\begin{equation}
    \ket{T(\vec{\boldsymbol{r}}), \boldsymbol{u}, \Delta} \equiv T(\vec{\boldsymbol{r}}) \ket{\boldsymbol{u}} + e^{i\Delta} \sqrt{1-|T(\vec{\boldsymbol{r}})|^2} \ket{\boldsymbol{u}_\perp},
\end{equation}
where $\Delta$ is given by Eq.~\eqref{Phase_condition}. The vector beam $\ket{T(\vec{\boldsymbol{r}}), \boldsymbol{u}, \Delta}$ is so constructed that the projection of $E_i(\vec{\boldsymbol{r}}) \ket{T(\vec{\boldsymbol{r}}), \boldsymbol{u}, \Delta}$ along the SoP $\ket{\boldsymbol{u}}$ yields $E_f(\vec{\boldsymbol{r}}) \ket{\boldsymbol{u}}$. All that remains to do is to identify whether there exists a d-plate that transforms $\ket{\boldsymbol{u}}$ to $\ket{T(\vec{\boldsymbol{r}}), \boldsymbol{u}, \Delta}$. If such a d-plate exists, then the light beam $E_i(\vec{\boldsymbol{r}}) \ket{\boldsymbol{u}}$ through it, followed by projection along the SoP $\ket{\boldsymbol{u}}$, results in the desired scalar beam $E_f(\vec{\boldsymbol{r}}) \ket{\boldsymbol{u}}$. This is true for any $\Delta$. We shall now find the right $\Delta$ such that the state $\ket{\boldsymbol{u}}$ transforms to $\ket{T(\vec{\boldsymbol{r}}), \boldsymbol{u}, \Delta}$ using a single d-plate. The desired $\Delta$ is, in principle, a function of $\ket{\boldsymbol{u}}$, and also of $\vec{\boldsymbol{r}}$ through $T(\vec{\boldsymbol{r}})$. If $\ket{\boldsymbol{u}}$ is a plane-polarized SoP, the expression for $\Delta$ simplifies to $\Delta = (2n-1)\pi + \Phi(\braket{\boldsymbol{l}|\boldsymbol{u}}) + \Phi(\braket{\boldsymbol{r}|\boldsymbol{u}})$. 
\par
Furthermore, with the choice of $\ket{\boldsymbol{u}} = \ket{\boldsymbol{h}}$, we have $\Delta=\pi/2$, and the vector beam $\ket{T(\vec{\boldsymbol{r}}), \boldsymbol{u}, \Delta}$ reduces to $\ket{T(\vec{\boldsymbol{r}})}$  of the form:
\begin{equation}
    \ket{T(\vec{\boldsymbol{r}})} = T(\vec{\boldsymbol{r}}) \ket{\boldsymbol{h}} + i\sqrt{1-|T(\vec{\boldsymbol{r}})|^2} \ket{\boldsymbol{v}}.
\label{VectorBeamT}
\end{equation}
The parameters $\Gamma(\vec{\boldsymbol{r}})$ and $\alpha(\vec{\boldsymbol{r}})$ of the d-plate that transforms $\ket{\boldsymbol{h}}$ to $\ket{T(\vec{\boldsymbol{r}})}$ can then be obtained from Eq.~\eqref{SingleWP_dplate} as: 
\begin{equation}
	\begin{aligned}
		\Gamma(\vec{\boldsymbol{r}}) &= 2\cos^{-1} (\textbf{re}(T(\vec{\boldsymbol{r}}))), \\
		\alpha(\vec{\boldsymbol{r}}) &= \frac{1}{2}\tan^{-1}\Bigg(\frac{\sqrt{1-|T(\vec{\boldsymbol{r}})|^2}}{\textbf{imag}(T(\vec{\boldsymbol{r}}))}\Bigg),
	\end{aligned}
	\label{ParametersforT}
\end{equation} 
where \textbf{re}($T(\vec{\boldsymbol{r}})$) and \textbf{imag}($T(\vec{\boldsymbol{r}})$) stand for the real and imaginary parts of the complex number $T(\vec{\boldsymbol{r}})$ at each $\vec{\boldsymbol{r}}$, respectively. From the above equation it is evident that if $T(\vec{\boldsymbol{r}})$ is real at some $\vec{\boldsymbol{r}}$, then the fast-axis orientation of the desired waveplate at that $\vec{\boldsymbol{r}}$ is $\pi/4$. Likewise, at all $\vec{\boldsymbol{r}}$ with $|T(\vec{\boldsymbol{r}})|=1$, the fast-axis will be oriented along the x-axis. 
\par
There is no loss of generality in considering transformations from scalar SoP $\ket{\boldsymbol{h}}$ to a vector beam $\ket{\boldsymbol{s}(\vec{\boldsymbol{r}})}$ of form Eq.~\eqref{VectorBeam}, since a transformation that takes $\ket{\boldsymbol{u}(\vec{\boldsymbol{r}})}$ to $\ket{\boldsymbol{w}(\vec{\boldsymbol{r}})}$, transforms $\ket{\boldsymbol{h}}$ to $\ket{\boldsymbol{s}(\vec{\boldsymbol{r}})}$ where $s_h(\vec{\boldsymbol{r}})=\bar{u}_h(\vec{\boldsymbol{r}}) w_h(\vec{\boldsymbol{r}}) + u_v(\vec{\boldsymbol{r}}) \bar{w}_v(\vec{\boldsymbol{r}})$ and $s_v(\vec{\boldsymbol{r}})=\bar{u}_h(\vec{\boldsymbol{r}}) w_v(\vec{\boldsymbol{r}}) - u_v(\vec{\boldsymbol{r}}) \bar{w}_h(\vec{\boldsymbol{r}})$. Here, the terms with subscripts $h$ and $v$ indicate the horizontal and vertical components of the corresponding vector beams.

\section{\label{Sec:Illustrations} Illustrations} 
\subsection{\label{Sec:Illustrations_plates} Representing SU(2) plate and waveplate}
Here we shall limit ourselves to cylindrical vector beams \cite{zhan2009cylindrical}, where the spatial variation is only along the azimuthal direction. The SU(2) transformation, which we now call as the SU(2) plate from now on, will then have its rotation angle $\Gamma$ and rotation axis $\boldsymbol{k}$ indicated as a vector in 3D with unit radius $(\theta, \varphi)$ as a function of the azimuthal angle $\phi$. So, we plot the normalized SU(2) parameters, $\frac{\Gamma(\phi)}{2\pi}$, $\frac{\theta(\phi)}{2\pi}$ and $\frac{\varphi(\phi)}{2\pi}$ as a polar plot in the colors orange, red, and green, respectively. As seen in Section~\ref{Sec:VBT}, the, SU(2) plates can be realized using single or multiple waveplates. These waveplates will also have their retardance $\Gamma$ and fast-axis orientations $\alpha$ as a function of the azimuthal angle $\phi$. We plot the normalized parameters $\frac{\Gamma(\phi)}{2\pi}$ and $\frac{\alpha(\phi)}{\pi}$ in the colors orange and blue respectively, as a function of the azimuthal angle $\phi$. We have also introduced another way of representing the SU(2) plate where the rotation angle $\Gamma$ is represented as a color-coded background and the rotation axis $\boldsymbol{k}$ is represented as ellipses. The rotation axis is related to the Stokes vector as $\boldsymbol{k} = -\hat{s}(\ket{\boldsymbol{s}})_i$, $i=x,y,z$. Each Stokes vector on the Poincar\'e sphere is associated with an SoP. The SoPs on the sphere are such that the poles include circular polarization states, the states on the equator are plane polarized, and the states at all other points are elliptically polarized with varying ellipticity, orientations and helicity. The components of polarization state are used to calculate the parameters of the ellipse, which determine the orientation of ellipse and hence the rotation axis. As we have seen in Section \ref{Sec:Theory}, an arbitrary SoP is defined as $\ket{\boldsymbol{s}}=s_h\ket{\boldsymbol{h}}+s_v\ket{\boldsymbol{v}}$ where $s_h$ and $s_v$ are the horizontal and vertical components of the SoP. The parameters of the ellipse are then given as:
\begin{equation}
    \begin{aligned}
        \chi&=\frac{1}{2}\sin^{-1}\big(\sin(2\alpha)\sin(\delta)\big) \\
        E&=|\tan(\chi)| \\
        \Phi&=\frac{\Phi(z)}{2}
        \label{Illustration_SU(2)plate}
    \end{aligned}
\end{equation}
where $\alpha=\tan^{-1}\bigg(\frac{|s_v|}{|s_h|}\bigg)$, $\delta=\Phi(s_v)-\Phi(s_h)$ and $z=\frac{s_h+is_v}{s_h-is_v}$. Here, $E$ and $\Phi$ are the ellipticity and orientation of the ellipse respectively. From Eq.~\eqref{Illustration_SU(2)plate} we get ellipticity $E=0$ for plane polarized light and orientation $\Phi=0,\pi/2,\pi/4$ and $3\pi/4$ for standard plane polarized light such as horizontal, vertical, diagonal and anti-diagonal respectively. For right and left circular polarization states we get $E=1$ and $\Phi=0$. The helicity is denoted by the color of the ellipse, with red color representing the points in the southern hemisphere and green color representing the points in the northern hemisphere, whereas, plane polarized light is indicated by black line.
\par
The illustrations depicting the evolution of a vector beam through an SU(2) plate and through its two and three waveplate equivalent is depicted in the appendix section. Now we give some illustrations of the SU(2) plate and d-plates  for the ideas discussed in Sections~\ref{Sec:VBT} and~\ref{Sec:CBS}. These examples illustrate how a diverse set of optical transformations can be viewed as specific instances of spatially varying SU(2) operations implemented by d-plates.

\subsection{Applications in quantum walks}
\subsubsection{Biased quantum step}

We now introduce a generalization of the quantum step given in Eq.~\eqref{stdquantumstep}, indicated by the symbol $\hat{T}(p,q;\boldsymbol{c},\boldsymbol{s})$ \cite{kadiri2024scouring}, where $p$ and $q$ are the step size, and $\boldsymbol{c}$ and $\boldsymbol{s}$ stand for the two SoPs $\ket{\boldsymbol{c}}$ and $\ket{\boldsymbol{s}}$, respectively. Unlike in the standard DTQW where $p$ and $q$ are step sizes of equal magnitude and opposite direction, here in the generalized quantum step, $p$ and $q$ can take any integer values. The action of the quantum step operator $\hat{T}(p,q;\boldsymbol{c},\boldsymbol{s})$ on the orthogonal pair of states $\ket{\boldsymbol{s};g}$ and $\ket{\boldsymbol{s}_\perp;g}$ is defined as:
\begin{equation}
    \begin{aligned}
        \hat{T}(p,q;\boldsymbol{c},\boldsymbol{s}) \ket{\boldsymbol{s};g} &= \ket{\boldsymbol{c};g+p} \\
        \hat{T}(p,q;\boldsymbol{c},\boldsymbol{s}) \ket{\boldsymbol{s}_\perp;g} &= \ket{\boldsymbol{c}_\perp;g+q}
    \end{aligned}
\label{GenQuantumStep1}
\end{equation}
The quantum step $\hat{T}(p,q;\boldsymbol{c},\boldsymbol{s})$ shifts the OAM by $p$ units if the SoP is $\ket{\boldsymbol{s}}$ and $q$ units if the SoP is $\ket{\boldsymbol{s}_\perp}$. In addition to the change in OAM, the quantum step also changes the SoP to $\ket{\boldsymbol{c}}$ in the former case, and to $\ket{\boldsymbol{c}_\perp}$ in the latter case. We don't assume any relationship between the SoPs $\ket{\boldsymbol{s}}$ and $\ket{\boldsymbol{c}}$: they need not be orthogonal like in q-plate, they could even be identical, or differ only by a global phase. 
The biased quantum step $\hat{T}(p,q;\boldsymbol{c},\boldsymbol{s})$ can be realized using two operators:
\begin{equation}
    \hat{T}(p,q;\boldsymbol{c},\boldsymbol{s}) = \hat{T}_{sy}\big(n; \boldsymbol{c}, \boldsymbol{s}\big) \hat{Q}_m 
\label{GenQuantumStep2}
\end{equation}
where $m=\frac{p+q}{2}$ and $n=\frac{p-q}{2}$. Here, the action of the operator $\hat{Q}_m$ follows from Eq.~\eqref{SPP} and the action of $\hat{T}_{sy}(n;\boldsymbol{c},\boldsymbol{s})$ is defined as:
\begin{equation}
    \begin{aligned}
    \hat{T}_{sy}(n;\boldsymbol{c},\boldsymbol{s}) \ket{\boldsymbol{s};g} &= \ket{\boldsymbol{c}; g+n} \\
    \hat{T}_{sy}(n;\boldsymbol{c},\boldsymbol{s}) \ket{\boldsymbol{s}_\perp;g} &= \ket{\boldsymbol{c}_\perp; g-n}
    \end{aligned}
\end{equation}
such that it moves the OAM $n$ steps forward if the SoP is $\ket{\boldsymbol{s}}$ and $n$ steps backward if the SoP is $\ket{\boldsymbol{s}_\perp}$, in addition to changing the SoP to $\ket{\boldsymbol{c}}$ and $\ket{\boldsymbol{c}_\perp}$ for the two cases respectively. Eq.~\eqref{GenQuantumStep2} can be validated by verifying the action of the two operators on the orthogonal states $\ket{\boldsymbol{s}; g}$ and $\ket{\boldsymbol{s_\perp}; g}$:
\begin{equation}
\begin{split}
    \ket{\boldsymbol{s}; g} \xrightarrow{\hat{Q}_m} \ket{\boldsymbol{s}; g+m} \xrightarrow{\hat{T}_{sy}(n, \boldsymbol{c}, \boldsymbol{s})} \ket{\boldsymbol{c};g+p} \\
    \ket{\boldsymbol{s}_\perp; g} \xrightarrow{\hat{Q}_m} \ket{\boldsymbol{s}_\perp; g+m} \xrightarrow{\hat{T}_{sy}(n, \boldsymbol{c}, \boldsymbol{s})} \ket{\boldsymbol{c}_\perp;g+q}
\end{split}
\end{equation}
which is the same as Eq.~\eqref{GenQuantumStep1}, thus validating the expression in Eq.~\eqref{GenQuantumStep2}.
\par
We now try to obtain the SU(2) transformation that implements the quantum step $\hat{T}(p,q;\boldsymbol{c},\boldsymbol{s})$. To achieve this, we find the SU(2) transformation of the quantum step $\hat{T}_{sy}(n;\boldsymbol{c},\boldsymbol{s})$. The parameters $\Gamma(\boldsymbol{\vec{r}})$ and $\boldsymbol{k}(\boldsymbol{\vec{r}})$ that implements the quantum step $\hat{T}_{sy}(n;\boldsymbol{c},\boldsymbol{s})$ are:
\begin{equation}
    \begin{aligned}
        \cos\frac{\Gamma(\boldsymbol{\vec{r}})}{2} &= |\braket{\boldsymbol{s}|\boldsymbol{c}}| \cos\big(n\phi + \Phi(\braket{\boldsymbol{s}|\boldsymbol{c}})\big) \\
        k_i(\boldsymbol{\vec{r}}) \sin\frac{\Gamma(\boldsymbol{\vec{r}})}{2} &= -|\braket{\boldsymbol{s}|\hat{\sigma}_i|\boldsymbol{c}}| \sin\big(n\phi + \Phi(\braket{\boldsymbol{s}|\hat{\sigma}_i|\boldsymbol{c}})\big) \\ & \text{for} \ i=x,y,z
    \end{aligned}
\label{SymQuantumStep_Param}
\end{equation}
Here $\Phi(c)$ denotes the phase of the complex number $c$. It is evident that if the SoPs $\ket{\boldsymbol{s}}$ and $\ket{\boldsymbol{c}}$ differ only by a global phase, that is, if $\ket{\boldsymbol{c}} = e^{i\Delta}\ket{\boldsymbol{s}}$, then the rotation angle varies linearly with the azimuthal angle as $\Gamma(\phi)=2(n\phi + \Delta)$, whereas the rotation axis is independent of the azimuthal angle, pointing in the negative direction of their (identical) Stokes vector, $\boldsymbol{k} = -\hat{s}(\ket{\boldsymbol{s}}) \equiv -\hat{s}(\ket{\boldsymbol{c}})$, defined in Eq.~\eqref{StokesVector}. On the other hand, if the two SoPs are distinct and differ in more than just a global phase, then the rotation axes $\boldsymbol{k}(\phi)$ vary with the azimuthal angle $\phi$, but is confined to a plane. The normal $\hat{\boldsymbol{n}}$ of the plane containing all the rotation axes is given by:
\begin{equation}
    \hat{\boldsymbol{n}} = \frac{\hat{s}(\ket{\boldsymbol{s}})-\hat{s}(\ket{\boldsymbol{c}})}{|\hat{s}(\ket{\boldsymbol{s}})-\hat{s}(\ket{\boldsymbol{c}})|},
\end{equation}
where $\hat{s}(\ket{\boldsymbol{s}})$ and $\hat{s}(\ket{\boldsymbol{c}})$ are the Stokes vectors of the SoPs $\ket{\boldsymbol{s}}$ and $\ket{\boldsymbol{c}}$ respectively. On the Poincar\'e sphere, this plane is a great circle, so located that the reflection of $\hat{s}(\ket{\boldsymbol{s}})$ about it yields $\hat{s}(\ket{\boldsymbol{c}})$. If the SoPs $\ket{\boldsymbol{s}}$ and $\ket{\boldsymbol{c}}$ are orthogonal, then it follows that the rotation angle is equal to $\pi$, at all azimuthal angles $\phi$. The quantum step $\hat{T}_{sy}(n;\boldsymbol{c},\boldsymbol{s})$ can be implemented using only a single d-plate, provided $k_z(\phi)=0 \ \forall \ \phi$. This is possible only if $\braket{\boldsymbol{s}|\sigma_z|\boldsymbol{c}}=0$. That is, when $\ket{\boldsymbol{s}}$ and $\ket{\boldsymbol{c}}$ are enantiogyres, which are states whose Stokes vectors are mirror reflections of each other about the equatorial plane of the Poincar\'e sphere. The normal $\hat{\boldsymbol{n}}$ in this case is along the $\hat{z}$ direction. 
\par
\begin{figure}[H]
    \centering    \includegraphics[width=1.0\linewidth]{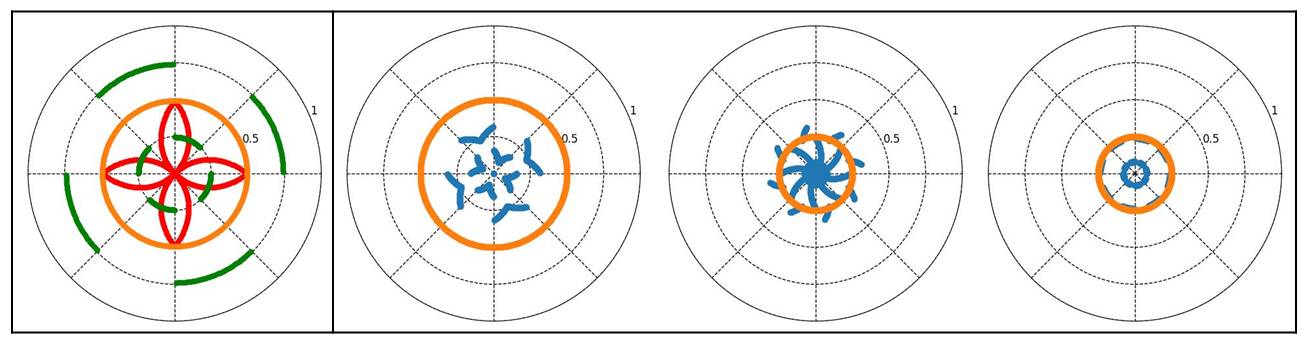} 
    \caption{Transformation corresponding to the quantum step $\hat{T}_{sy}\big(4,\boldsymbol{h},\boldsymbol{v}\big)$. The plot towards the left hand side represents the parameters of SU(2) plate with orange line indicating the rotation angle $\Gamma$, the red and green lines indicating $\theta$ and $\varphi$ respectively, of the rotation axis. The plots towards the right hand side represents the parameters of $\hat{H}\hat{Q}_1\hat{Q}_2$, the three waveplate equivalent of the SU(2) plate. Here, the orange line represents retardance $\Gamma$ and blue line represents fast-axis orientation $\alpha$.}
    \label{Biased_QW_Parameters}
\end{figure}

For illustration, we consider the quantum step $\hat{T}(3,-5,\boldsymbol{h},\boldsymbol{v})$. The above quantum step can be written in the form of Eq.~\eqref{GenQuantumStep2} with $m=-1$ and $n=4$ as:
\begin{equation}
    \hat{T}(3,-5,\boldsymbol{h},\boldsymbol{v}) = \hat{T}_{sy}\big(4,\boldsymbol{h},\boldsymbol{v}\big) \hat{Q}_{-1}
    \label{Biased_QW_Illustration}
\end{equation}
The parameters of the SU(2) transformation to implement the step $\hat{T}_{sy}\big(4,\boldsymbol{h},\boldsymbol{v}\big)$ is determined from Eq.~\eqref{SymQuantumStep_Param}. The term $|\braket{\boldsymbol{s}|\boldsymbol{c}}|$ in this case becomes $|\braket{\boldsymbol{v}|\boldsymbol{h}}|=0$ since the SoPs $\ket{\boldsymbol{v}}$ and $\ket{\boldsymbol{h}}$ are orthogonal to each other. Hence, the rotation angle becomes $\Gamma=\pi$. The SU(2) plate to implement the quantum step $\hat{T}_{sy}\big(4,\boldsymbol{h},\boldsymbol{v}\big)$ in the above example and its three waveplate equivalent are represented as polar plots in Fig.~\ref{Biased_QW_Parameters}. 
The action of the quantum step $\hat{T}(3,-5,\boldsymbol{h},\boldsymbol{v})$ involves the action of a SPP followed by $\hat{T}_{sy}\big(4,\boldsymbol{h},\boldsymbol{v}\big)$ as given in Eq.~\eqref{Biased_QW_Illustration}. The latter can be realized using a HWP and two QWPs as depicted in Fig.~\ref{Biased_QW_Parameters}. An example showing the action of the quantum step $\hat{T}(3,-5,\boldsymbol{h},\boldsymbol{v})$ on a state is shown in Figs.~\ref{Biased_QW_evolution_of_(v;0)_HV_components} and~\ref{Biased_QW_evolution_of_(-h;0)_HV_components} respectively of Appendix~\ref{Sec:Appendix_BiasedQS}. 

\subsubsection{\label{Sec:InstantaneousQW}Instantaneous quantum walks}
\begin{figure}[H]
    \centering
    \includegraphics[width=1.0\linewidth]{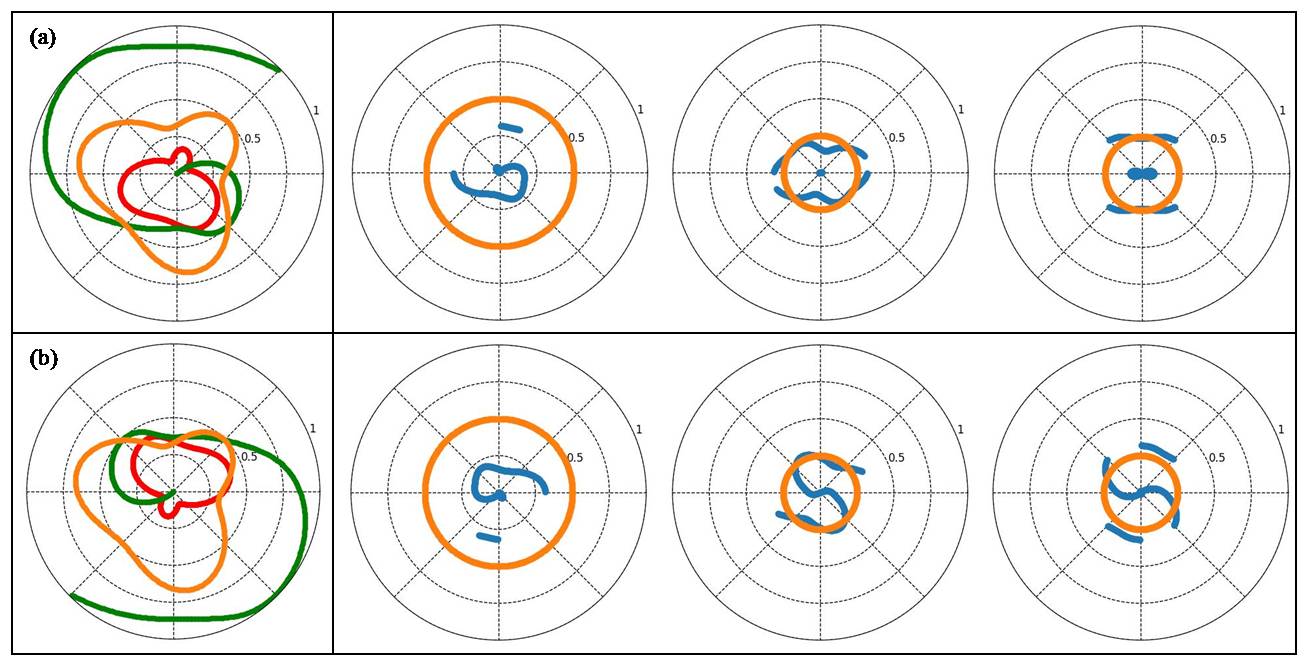} 
    \caption{(a), (b) Transformations corresponding to $\hat{S}_1$ and $\hat{S}_2$  respectively, given in Eq.~\eqref{UtoW_transformation}. The plots towards the left hand side represent the parameters of the SU(2) plate with orange line indicating the rotation angle $\Gamma$, the red and green lines indicating $\theta$ and $\varphi$ respectively, of the rotation axis. The plots towards the right hand side represent the parameters of $\hat{H}\hat{Q}_1\hat{Q}_2$, the three waveplate equivalent of the SU(2) plate with orange line representing retardance $\Gamma$ and blue line representing fast axis orientation $\alpha$.}
    \label{InstantQW}
\end{figure}

We define an instantaneous quantum walk as a single engineered unitary transformation acting on the composite Hilbert space containing the SAM and OAM states, that directly maps an initial composite state to a final state that would otherwise be obtained after multiple steps of a DTQW. The instantaneous quantum walk is a compact physical realization strategy of multi-step walk dynamics via a single composite unitary operation. Here we show the transformation between two such composite states in the optical realization. Consider two vector beams:
\begin{equation}
    \begin{aligned}
        \ket{\boldsymbol{U}} &= \frac{1}{\sqrt{2}} \big(\ket{\boldsymbol{l};1} + \ket{\boldsymbol{r};-1}\big) \\
        \ket{\boldsymbol{W}} &= \frac{1}{\sqrt{2}} \big(\ket{\boldsymbol{d};0} + \ket{\boldsymbol{a};2}\big)
        \label{InstantQW_example}
    \end{aligned}
\end{equation}
We now discuss the two SU(2) transformations that take the state $\ket{\boldsymbol{U}}$ to the state $\ket{\boldsymbol{W}}$ and vice-versa.
\begin{equation}
    \begin{aligned}
        \hat{S}_1 \ket{\boldsymbol{U}} = \ket{\boldsymbol{W}} \\
        \hat{S}_2 \ket{\boldsymbol{W}} = \ket{\boldsymbol{U}}
    \end{aligned}
\label{UtoW_transformation}
\end{equation}
The parameters of the SU(2) plates for $\hat{S}_1$ and $\hat{S}_2$ are obtained from Eq.~\eqref{SU2Parameters}. These parameters and its equivalent physical implementation with three waveplates are represented as polar plots in Fig.~\ref{InstantQW}. 
The states $\ket{\boldsymbol{U}}$ and $\ket{\boldsymbol{W}}$ as it evolves through the waveplates depicted in Fig.~\ref{InstantQW} are represented in Figs.~\ref{Beam_InstantQW_input_U_evolution} and \ref{Beam_InstantQW_input_W_evolution} respectively of Appendix~\ref{Sec:Appendix_Instant_QW}.

\subsection{Full Poincar\'e beams}
In Section \ref{Sec:Theory} we have introduced the classification of light beam as scalar and vector beams. A scalar beam has the same SoP across the beam cross-section. It can be represented as a single point on the Poincar\'e sphere and can be denoted as,
\begin{equation}
    \ket{\theta, \varphi} = \cos{\frac{\theta}{2}}\ket{\boldsymbol{l}} + e^{i(\varphi-\frac{\pi}{2})} \sin{\frac{\theta}{2}}\ket{\boldsymbol{r}}
    \label{pol_state}
\end{equation}
where $\theta$ and $\varphi$ are the coordinates on the sphere taking values between $(0, \pi)$ and $(0, 2\pi)$ respectively. A vector beam has spatially varying SoPs and are represented by a collection of points on the Poincar\'e sphere. As a special case, the collection of points for a vector beam can be such that it contains all possible points on the Poincar\'e sphere. The transverse profile of such a beam would have all possible SoPs and are called Full Poincar\'e Beams (FPB) \cite{beckley2010full}. This kind of mapping where each point on the sphere is mapped onto a plane can be done using Lambert Azimuthal Projection (LAP) \cite{lopez2019overall,snyder1987map} which is an equal area mapping. The polar and azimuthal coordinates $(\theta, \varphi)$ of the Poincar\'e sphere are mapped onto a disk with radial and azimuthal coordinates $(r, \phi)$, representing the transverse profile of the beam. A state of an LAP FPB can be written as $\ket{\theta,\varphi}$ given in Eq.~\eqref{pol_state} with:
\begin{equation}
    \theta=2\cos^{-1}\bigg(\frac{r}{r_0}\bigg) \ \text{and}\ \varphi= \phi 
\end{equation}
where $r_0$ is the radius of the disk and is proportional to the radius of the sphere. For the above mapping, the polarization state at the point of contact of the sphere with the disk is mapped onto the centre of the disk, while the SoP corresponding to its antipodal point is mapped on the circumference of the disk. So, the 
polarization states at $\theta=\pi$ and $\theta=0$ are placed at the centre and periphery of the disk respectively. The linearly polarized states at the equator with $\theta=\frac{\pi}{2}$ are mapped onto the disk at a radius of $r=\frac{r_0}{\sqrt{2}}$.
\par
Another kind of mapping of an FPB onto a plane is that of optical skyrmions \cite{donati2016twist, shen2024optical}. The state of an FPB for skyrmion mapping can then be obtained from Eq.~\eqref{pol_state}: 
\begin{equation}
    \theta=2\tan^{-1}\bigg(\frac{r_0-r}{r}\bigg) \ \text{and}\ \varphi= \phi 
\end{equation}
The skyrmion mapping places the circular polarization states corresponding to $\theta=\pi$ and $\theta=0$ at the centre and periphery of the disk respectively, while the plane polarized states at $\theta=\frac{\pi}{2}$ are placed on the disk at a radius of $r=\frac{r_0}{2}$. 
\par
\begin{figure}[H]
    \centering
    \includegraphics[width=1.0\linewidth]{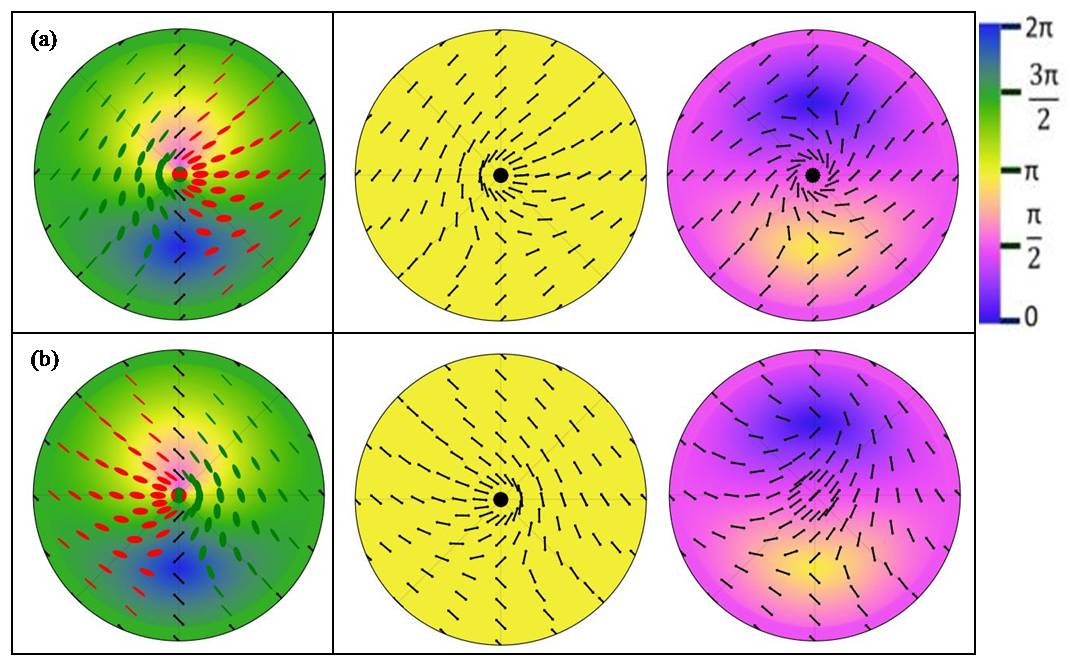}
    \caption{The d-plate and waveplate parameters corresponding to the transformation between a scalar SoP and an FPB. The plot towards left hand side shows the d-plate parameters and the plots towards right hand side shows the two waveplate equivalent of the same. (a) Transformation from a scalar beam with horizontal SoP to skyrmion FPB with right-circular polarization state at the center, (b) Transformation from skyrmion FPB with right-circular polarization at the center to a scalar beam with horizontal SoP. The plots describes the fast axis orientation and the retardance at each point of the plate where, retardance is represented as a color coded background.}
    \label{FPB_transformation_paramaters}
\end{figure}

As the SoP of an FPB is varying across its transverse profile, the waveplate required to generate an FPB in turn has to be inhomogeneous, that is, its retardance and fast-axis orientation has to vary spatially. The parameters of the d-plate that generates an FPB is derived from Eq.~\eqref{SingleWP_dplate}. The d-plate can also transform one FPB to another. Illustrations of transformations between arbitrary scalar SoP and LAP FPB is shown in Ref. \cite{radhakrishna2022polarimetric}. Here, we show the transformation between a horizontally polarized scalar beam and a skyrmion FPB with right circular polarization at its centre. The d-plate parameters for the above transformation and its equivalent two waveplate implementation is shown in Fig.~\ref{FPB_transformation_paramaters}. 
The evolution of light beam through these plates is shown in Appendix~\ref{Sec:Appendix_FPB}.

\subsection{Complex beam shaping}
\subsubsection{\label{Sec:Anglestates} Angle states
}
As an illustration of complex beam shaping using d-plates, we will now attempt to generate arbitrary state $\ket{c}_o$ of the OAM space:
\begin{equation}
    \ket{c}_o = \sum_{m=b}^e c_m\ket{m}_o
\label{OAMstate}
\end{equation}
where $c_m, \ m=b,\dots,e$ are complex numbers such that $\sum_{m=b}^e|c_m|^2=1$. Here $b$ and $e \geq b$ are integers. 
\par
Given any such state $\ket{c}_o$ in the OAM space, we will associate a complex function $c(\phi)$ as,
\begin{equation}
    c(\phi)=\sum_{m=b}^e c_m e^{im\phi}
\end{equation}
To accomplish this task, we seek the vector beam $\ket{\tilde{c}(\phi)}$ defined in the form of Eq.~\eqref{VectorBeamT} as:
\begin{equation}
    \ket{\tilde{c}(\phi)} = \tilde{c}(\phi) \ket{\boldsymbol{h}} + i\sqrt{1-|\tilde{c}(\phi)|^2} \ket{\boldsymbol{v}}
\label{VectorBeamC}
\end{equation}
where $\tilde{c}(\phi)$ is $c(\phi)$ normalized to 1: $\tilde{c}(\phi)=\frac{c(\phi)}{\max_\phi(|c(\phi)|)}$. The projection of this vector beam $\ket{\tilde{c}(\phi)}$ along the $\ket{\boldsymbol{h}}$ direction yields $\tilde{c}(\phi) \ket{\boldsymbol{h}}$, where the function $\tilde{c}(\phi)$ is the vector $\ket{c}_o$ in the OAM space, upto a normalization factor. Here, the state $\ket{c}_o$ is thus generated from a vector beam. 
\par
\begin{figure}[H]
    \centering
    \includegraphics[width=1.0\linewidth]{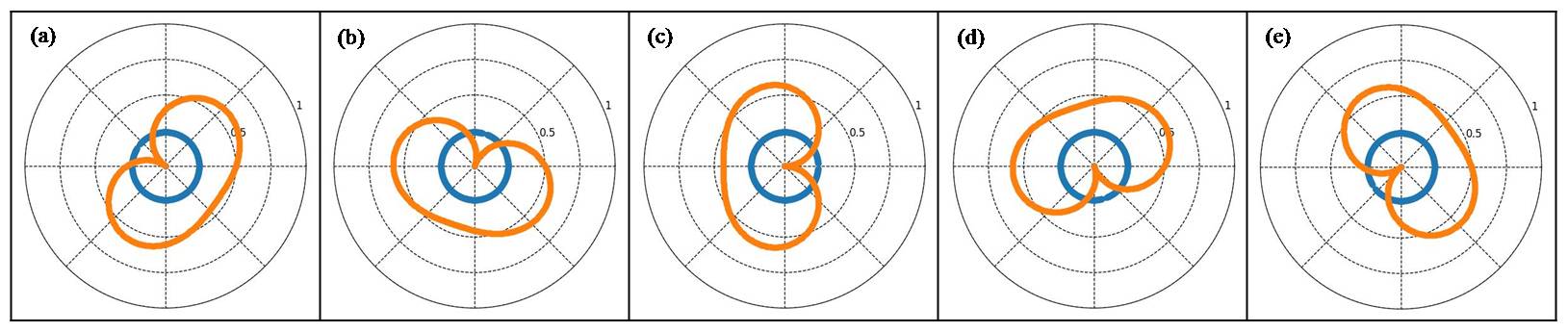}
    \caption{(a-e) The d-plates required for generating the five angle states $\ket{2,j}$, for $j=-2, \dots, 2$ respectively, where the orange line indicates retardance $\Gamma$ and the blue line indicates fast-axis orientation $\alpha$.}
    \label{Anglestates_Param}
\end{figure}

As a first demonstration of this idea, we seek to construct the so called angle states \cite{mirhosseini2015high,yao2006fourier}. These angle states are parameterized with two integers $N$ and $j$, and are defined as:
\begin{equation}
    \ket{N,j}_o = \frac{1}{\sqrt{2N+1}} \sum_{m=-N}^N e^{i\frac{2\pi mj}{2N+1}} \ket{m}_o
\end{equation}
The angle state $\ket{N,j}_o$ is such that its projection along each of the OAM $m$ is identical in magnitude for all $m=-N,\dots,N$. For a given $N$, valid $j$ are from $-N$ to $N$. Each of these $2N+1$ angle states makes equal angle with each of the $2N+1$ OAM eigen states $\ket{m}_o,\ m=-N,\dots,N$, and vice-versa. To generate the above angle state, we construct a vector beam of the form of Eq.~\eqref{VectorBeamC}, with $\tilde{c}(\phi)=\frac{1}{2N+1}\Big(1+2\sum_{m=1}^N \cos\big(m\phi + \frac{2\pi mj}{2N+1}\big)\Big)$. The retardance and fast axis orientation of the d-plate that generates these angle states are derived from Eq.~\eqref{ParametersforT} as:
\begin{equation}
    \begin{aligned}
        \alpha(\phi)&=\frac{\pi}{4}; \\ \Gamma(\phi)&=2\cos^{-1}\bigg[\frac{1}{2N+1}\bigg(1+\sum_{m=1}^N s_m(\phi)\bigg)\bigg]
    \end{aligned}
\end{equation}
where $s_m(\phi)$ is given by $s_m(\phi)=2 \cos\big(m\phi + \frac{2\pi mj}{2N+1}\big)$. An illustration of these d-plates for $N=2$ is depicted in Fig.~\ref{Anglestates_Param}.
The vector beams emerging at the exit plane of these plates and its transverse plane intensity profiles, for horizontal input is represented in Appendix~\ref{Sec:Appendix_anglestates}.

\subsection{Quantum Maps}
The action of a d-plate may also be interpreted as a quantum map on the orbital angular momentum (OAM) degree of freedom, with polarization serving as an ancilla subsystem. \par
The Kraus form of this map can be expressed as:
\begin{equation}
	\mathcal{E}(\rho_o) =  \hat{K}_h \rho_o \hat{K}_h^\dagger +\hat{K}_v \rho_o \hat{K}_v^\dagger
\end{equation}
where $\rho_o$ is an arbitrary density matrix in the OAM space, and $\hat{K}_h$ and $\hat{K}_v$ are the Kraus operators. 
with their action on the OAM basis vectors $|m\rangle_o$ is
\begin{equation}
\begin{aligned}
	\hat K_h |m\rangle_o
	&=
	\sum_n h_n |m+n\rangle_o,
	\\
	\hat K_v |m\rangle_o
	&=
	\sum_n v_n |m+n\rangle_o, 
\end{aligned}
\end{equation}
where the coefficients $h_n$ and $v_n$ are given as:
\begin{equation}
\begin{aligned}
	h_n
	&=
	\frac{1}{2\pi}
	\int_0^{2\pi}
	\langle h|\hat W_{\Gamma(\phi)}(\alpha(\phi))|h\rangle e^{-in\phi}d\phi,
	\\
	v_n
	&=
	\frac{1}{2\pi}
	\int_0^{2\pi}
	\langle v|\hat W_{\Gamma(\phi)}(\alpha(\phi))|h\rangle e^{-in\phi}d\phi.
\end{aligned}
\end{equation}
Equivalently, the matrix elements of the Kraus operators
are
\begin{equation}
\begin{aligned}
	{}_o\langle m'|
	\hat K_h
	|m\rangle_o
	&=
	h_{m'-m},
	\\
	{}_o\langle m'|
	\hat K_v
	|m\rangle_o
	&=
	v_{m'-m}.
	\label{eq:fourier_kraus}
\end{aligned}
\end{equation}
Since the matrix elements in Eq.~(\ref{eq:fourier_kraus}) depend only on the difference $(m'-m)$, the induced
Kraus operators assume convolutional form in the OAM basis. 
The map is completely positive by construction. Trace preservation follows directly from the pointwise unitarity of the waveplate operator $\hat{W}_{\Gamma(\phi)}(\alpha(\phi))$, which guarantees $|\langle h|\hat{W}|h\rangle|^2 + |\langle v|\hat{W}|h\rangle|^2 = 1$ at every azimuth $\phi$. By Parseval's theorem, this yields $\hat{K}_h^\dagger \hat{K}_h + \hat{K}_v^\dagger \hat{K}_v = \hat{I}_{\text{OAM}}$, ensuring the channel is strictly a Completely Positive Trace Preserving (CPTP) map \cite{Holevo2019}.

\begin{figure}[H]
    \centering
    \includegraphics[width=1.0\linewidth]{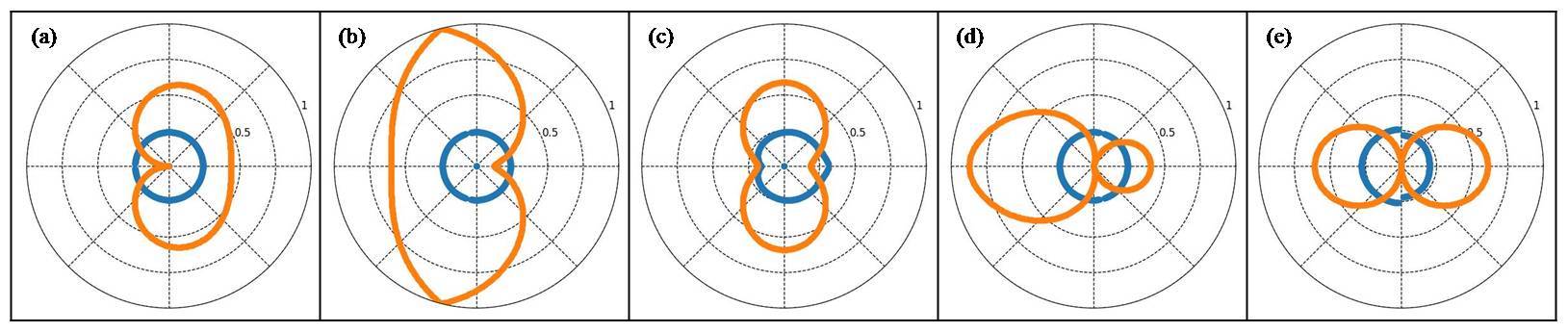}
    \caption{(a-e) The d-plates required for generating the five OAM states $\ket{o_i}$, for $i=1, \dots, 5$ of Eq.~\eqref{OAMStates_illus} respectively, where the orange line indicates retardance $\Gamma$ and the blue line indicates fast-axis orientation $\alpha$.}
    \label{OAMstates_Param}
\end{figure}

Consequently, the present framework realizes a structured class of translation-invariant quantum maps on the OAM degree of freedom through direct engineering of the azimuthally varying retardance and fast-axis orientation of a single d-plate.
As an illustration of generation of OAM states using the d-plate consider the following five states on the OAM space, of the form Eq.~\eqref{OAMstate} as:
\begin{equation}
    \begin{aligned}
        \ket{o_1} &= \frac{1}{\sqrt{5}} \big(\ket{-2} - \ket{-1} + \ket{0} - \ket{1} + \ket{2}\big) \\
        \ket{o_2} &= \frac{1}{\sqrt{5}} \big(\ket{-2} + \ket{-1} - \ket{0} + \ket{1} + \ket{2}\big) \\
        \ket{o_3} &= \frac{1}{\sqrt{5}} \big(\ket{-2} -i \ket{-1} + \ket{0} -i \ket{1} + \ket{2}\big) \\
        \ket{o_4} &= \frac{1}{\sqrt{5}} \big(-\ket{-2} + \ket{-1} + \ket{0} +  \ket{1} - \ket{2}\big) \\
        \ket{o_5} &= \frac{1}{\sqrt{5}} \big(-\ket{-2} +i \ket{-1} + \ket{0} +i \ket{1} - \ket{2}\big)
    \end{aligned}
\label{OAMStates_illus}
\end{equation}
The d-plates required to generate the five states given in the above equation from a horizontally polarized scalar beam is depicted in Fig.~\ref{OAMstates_Param}.
The vector beams emerging at the exit plane of these plates and its transverse plane intensity profiles, for horizontal input is represented in Appendix~\ref{Sec:Appendix_Quantummaps}.

\section{Conclusion}
\label{Sec:Conclusion}
To conclude, in this paper we have answered the following question: given two vector beams, whether they can be transformed from one to another using single waveplate? Our definition of the state of polarization includes the global phase factor also, and we desire the waveplate to transform from one to the other exactly, not up to a phase factor. Given two vector beams that satisfy this condition, we have identified a single waveplate that implements the transformation. We have also shown the implementation of its equivalent using two waveplates and three waveplates. In the second part, we have explored that in the spin-orbital angular momentum implementation of quantum walks it is possible to realize a quantum step that is more generalized than what is possible with a single q-plate. This generalization puts a generic elliptical polarization state of light on par with circular polarizations in inducing changes in the orbital angular momentum content of light. It has been shown that a time-independent, position-independent quantum walk comprising of generalized quantum steps can be realized using just a pair of inhomogeneous waveplates. Furthermore, we have examined the possibility of complex beam shaping using only one doubly-inhomogeneous waveplate. We have shown that it is possible to generate any desired amplitude distribution and wavefront profile from a scalar Gaussian light beam, using an appropriate d-plate and standard polarizer. We have shown that the action of certain class of doubly-inhomogeneous waveplates can be interpreted as the physical realization of the completely positive trace preserving maps with the polarization serving as an ancilla degree of freedom. All our results, though cast in the language of waveplates, apply equally well to even birefringent metasurfaces.

\begin{acknowledgments}
	GGT acknowledges the support of Indira Gandhi Centre for Atomic Research and the Homi Bhabha National Institute (HBNI), Mumbai, through the award of Research Fellowship. 
\end{acknowledgments}

\appendix
\section{\label{Sec:Illustrations_VectorBeam} Representing vector beam}

We depict the vector beam $\ket{\boldsymbol{s}(\phi)}$ in terms of its horizontal and vertical components $s_h(\phi)$ and $s_v(\phi)$. We represent these complex numbers in two different panels, plotting the intensities in the horizontal and vertical components, $|s_h(\phi)|^2$ and $|s_v(\phi)|^2$ as polar plots indicated by a black line, and the corresponding phases $\Phi(s_h(\phi))$ and $\Phi(s_v(\phi))$ as color-coded backgrounds of these polar plots with blue, pink, and yellow indicating the phases $0, \pi/2$ and $\pi$ respectively. The horizontal and vertical components of the vector beam is also represented by its transverse plane intensity profiles. We have also represented the transverse profile of Full Poincar\'e beam as polarization ellipses. The ellipticity and orientation of the ellipse is determined from the Stokes vector of the polarization state which are related as given in Eq.~\eqref{StokesVector}. So, $E$ an $\Phi$ are then given as:
\begin{equation}
    \begin{aligned}
        E&=||S_z|| \\
        \Phi&=\frac{1}{2}\tan^{-1}\bigg(\frac{S_y}{S_x}\bigg)
    \end{aligned}
    \label{Illustration_Vectorbeam}
\end{equation}
where $S_x,S_y,S_z$ are components of the Stokes vector $\boldsymbol{S}=(S_x,S_y,S_z)$. From Eq.~\eqref{Illustration_Vectorbeam},  the ellipticity for plane polarized light is determined to be $E=0$ and for standard plane polarized light such as horizontal, vertical, diagonal and anti-diagonal, the orientation is $\Phi=0,\pi/2,\pi/4$ and $3\pi/4$ respectively. For right and left circular polarization states, $E=1$ and $\Phi=0$. 

\section{\label{Sec:Appendix_BiasedQS} Biased quantum step}
We show the action of the quantum step $\hat{T}(3,-5;\boldsymbol{h},\boldsymbol{v})$ on a state $\ket{\boldsymbol{v};0}$ and its orthogonal state $-\ket{\boldsymbol{h};0}$. The evolution of these states as it passes through the plates $\hat{Q}_{-1}$ and $\hat{H}\hat{Q}_1\hat{Q}_2$ depicted in Fig.~\ref{Biased_QW_Parameters} are shown in the Figs.~\ref{Biased_QW_evolution_of_(v;0)_HV_components} and~\ref{Biased_QW_evolution_of_(-h;0)_HV_components} below. The top and bottom rows of both the figures, represents the projection of the emerging beam along the horizontal and vertical polarization states.
\begin{figure}[H]
    \centering
    \includegraphics[width=1.0\linewidth]{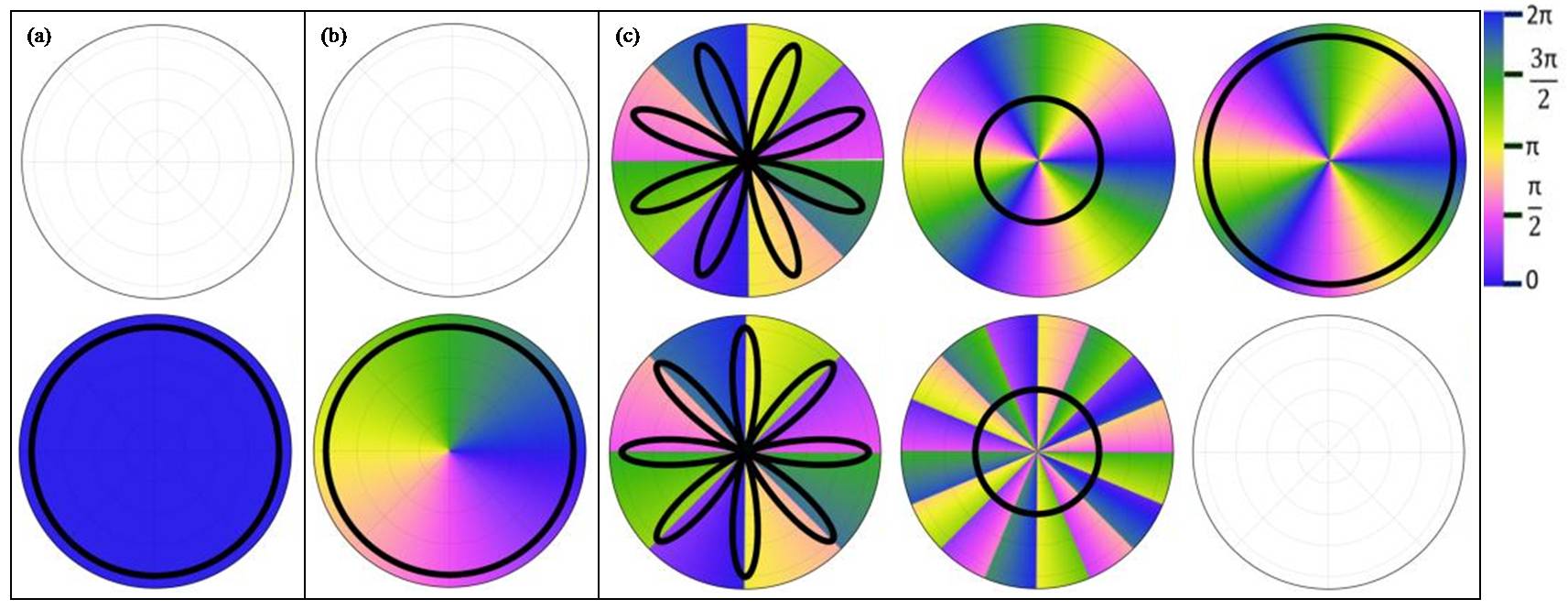} 
    \caption{Evolution of the beam with vertical polarization and zero OAM, $\ket{\boldsymbol{v};0}$ as it emerges from the plates $\hat{Q}_{-1}$ and $\hat{H}\hat{Q}_1\hat{Q}_2$. (a) Beam corresponding to the initial state $\ket{\boldsymbol{v};0}$, (b) and (c) corresponds to the output beam from $\hat{Q}_{-1}$ and $\hat{H}\hat{Q}_1\hat{Q}_2$ respectively.}
    \label{Biased_QW_evolution_of_(v;0)_HV_components}
\end{figure}

\begin{figure}[H]
    \centering
    \includegraphics[width=1.0\linewidth]{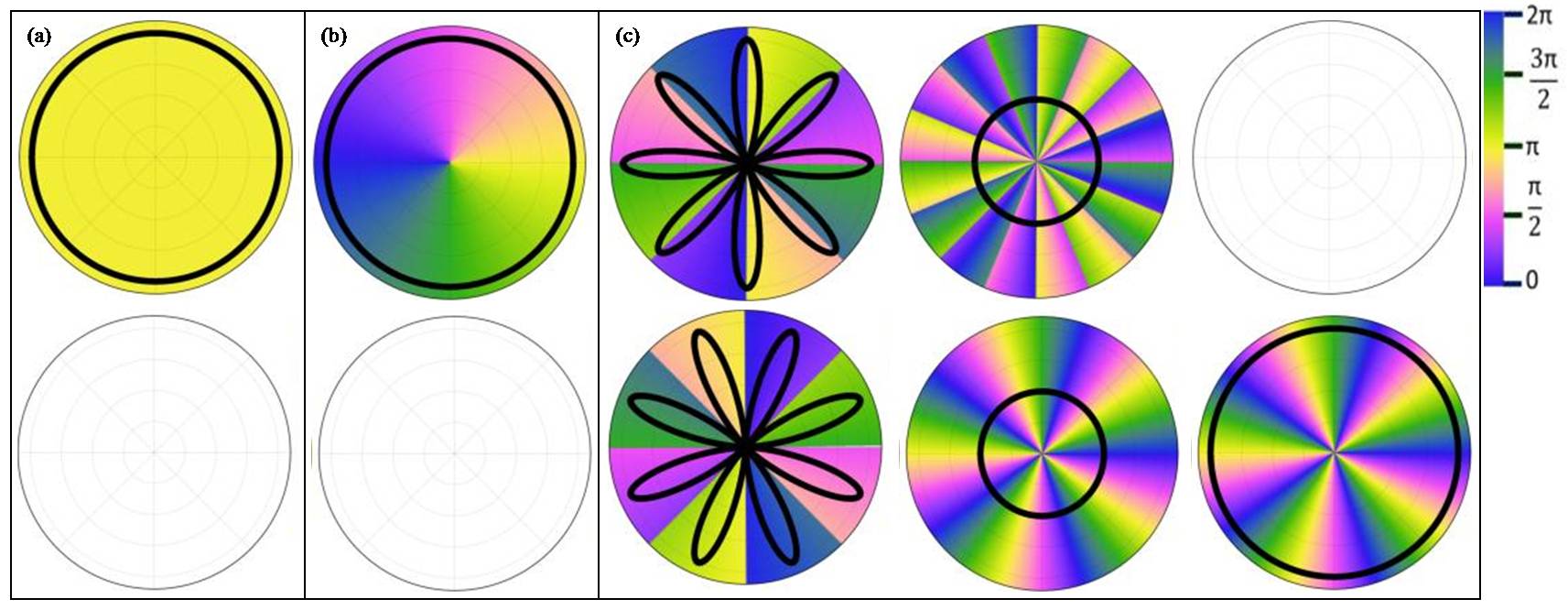} 
    \caption{Evolution of the beam with horizontal polarization, zero OAM and a global phase of $\pi$, $-\ket{\boldsymbol{h};0}$ as it emerges from the plates $\hat{Q}_{-1}$ and $\hat{H}\hat{Q}_1\hat{Q}_2$. (a) Beam corresponding to the initial state $-\ket{\boldsymbol{h};0}$, (b) and (c) corresponds to the output beam from $\hat{Q}_{-1}$ and $\hat{H}\hat{Q}_1\hat{Q}_2$ respectively.}
    \label{Biased_QW_evolution_of_(-h;0)_HV_components}
\end{figure}

It is to be noted in the above figures that the plots with no black line and white background indicates zero amplitude and hence absence of the respective component.

\section{\label{Sec:Appendix_Instant_QW} Instantaneous quantum walk}
\begin{figure}[H]
    \centering
    \includegraphics[width=1.0\linewidth]{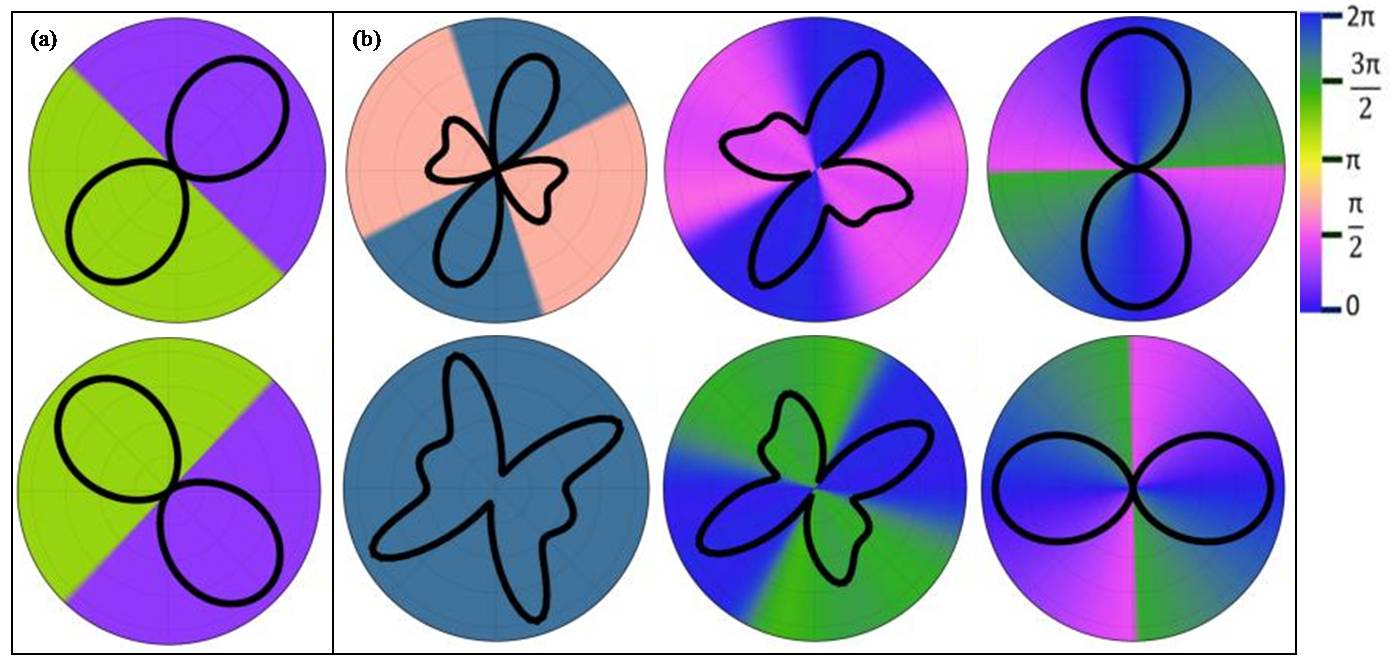}
    \caption{Evolution of the state $\ket{\boldsymbol{U}}$ as it passes through the waveplates $\hat{H}\hat{Q}_1\hat{Q}_2$. (a) Vector beam represented by the state $\ket{\boldsymbol{U}}$, (b) vector beam emerging through the set of waveplates shown in Fig.~\ref{InstantQW}(a).} 
    \label{Beam_InstantQW_input_U_evolution}
\end{figure}

\begin{figure}[H]
    \centering
    \includegraphics[width=1.0\linewidth]{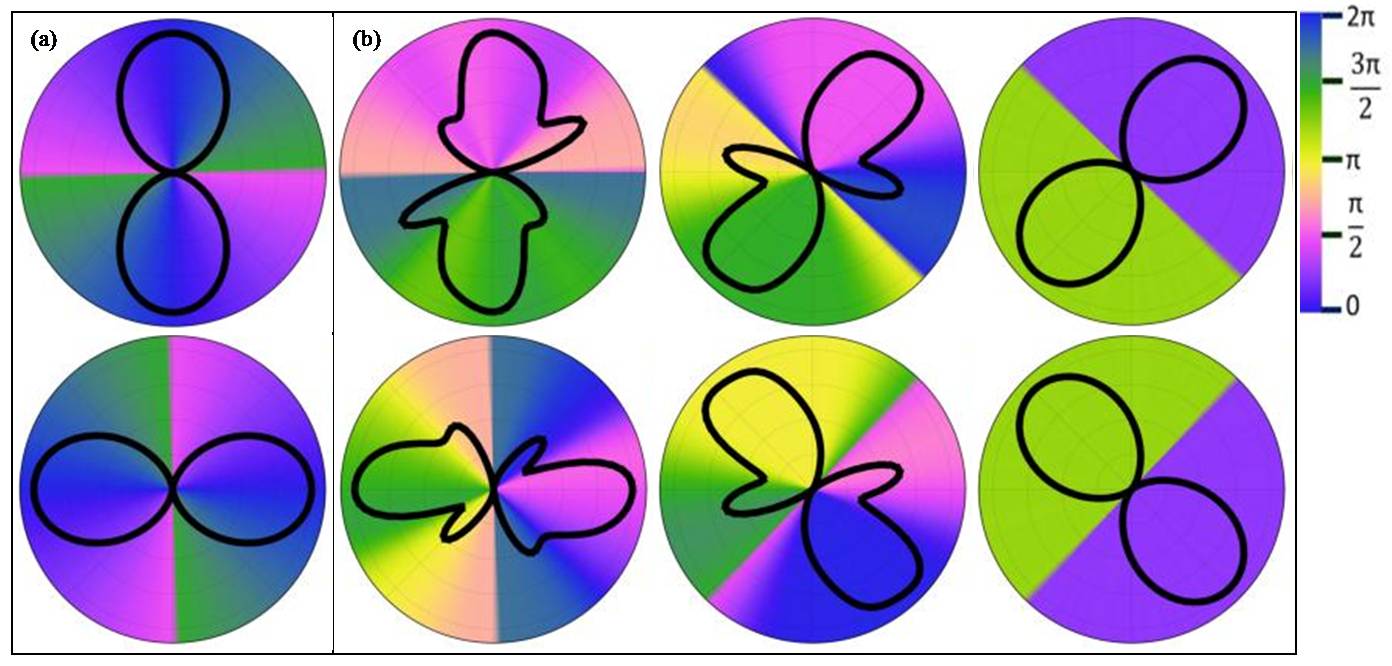}
    \caption{Evolution of the state $\ket{\boldsymbol{W}}$ as it passes through the waveplates $\hat{H}\hat{Q}_1\hat{Q}_2$. (a) Vector beam represented by the state $\ket{\boldsymbol{W}}$, (b) vector beam emerging through the set of waveplates shown in Fig.~\ref{InstantQW}(b).}
    \label{Beam_InstantQW_input_W_evolution}
\end{figure}

An illustration of the example given in Section~\ref{Sec:InstantaneousQW}. The evolution of the states $\ket{\boldsymbol{U}}$ and $\ket{\boldsymbol{W}}$ given in Eq.~\eqref{InstantQW_example} as it passes through the waveplates depicted in Fig.~\ref{InstantQW} is shown in Figs.~\ref{Beam_InstantQW_input_U_evolution} and~\ref{Beam_InstantQW_input_W_evolution} respectively. The top and bottom rows of both the figures, represents horizontal and vertical components of the emerging vector beam.

\section{\label{Sec:Appendix_FPB} Full Poincar\'e beams}
We show the illustration of transformation between a horizontally polarized scalar beam and a skyrmion FPB with right circular polarization at the center and vice-versa as these beams emerge from the plates depicted in Fig.~\ref{FPB_transformation_paramaters}. 
\begin{figure}[H]
    \centering
    \includegraphics[width=1.0\linewidth]{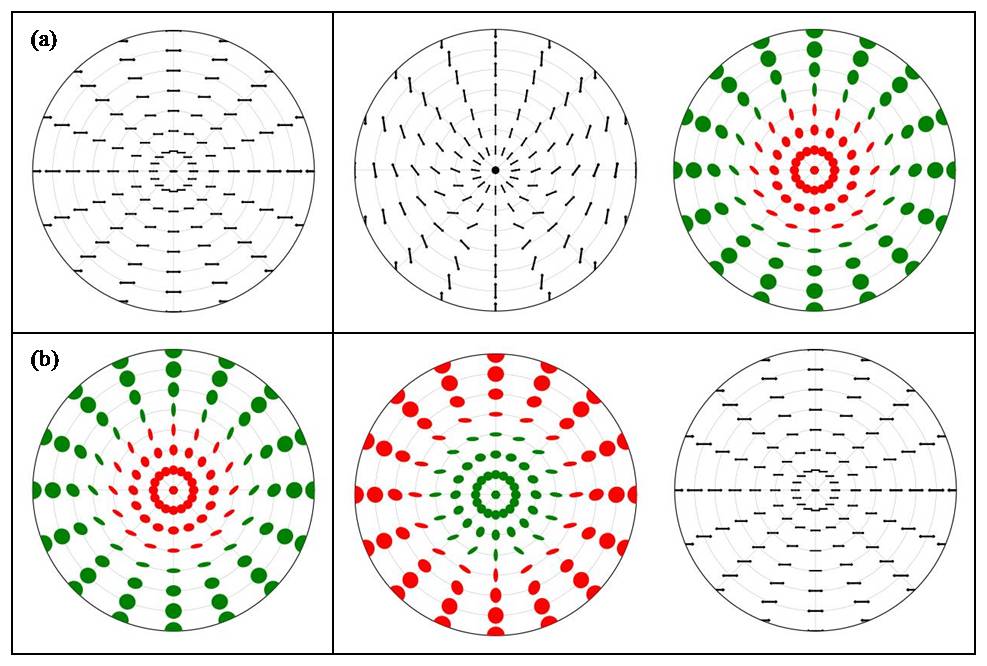}
    \caption{Evolution of an incident light beam through d-plate and waveplates. The plots towards left hand side shows the polarization pattern of the incident beam and the plots towards right hand side shows the evolution of the incident beam as it passes through the set of two waveplates shown in Fig.~\ref{FPB_transformation_paramaters}. (a) Evolution of a scalar beam with horizontal polarization, (b) Evolution of a skyrmion FPB with right circular polarization at the center.}
    \label{Beam_evolution_LAP_and_skyrmion}
\end{figure}
The ellipses represent the polarization states mapped onto a disk with circles indicating circular polarization and lines indicating plane polarization. All other ellipses with varying ellipticity and orientations indicate the SoPs on the Poincar\'e sphere as discussed in Section~\ref{Sec:Illustrations_VectorBeam}. The color of the ellipse indicates the part of hemisphere that it occupies, with red and green color representing the southern and northern hemispheres respectively.

\section{\label{Sec:Appendix_anglestates} Angle states}
\begin{figure}[H]
    \centering
    \includegraphics[width=1.0\linewidth]{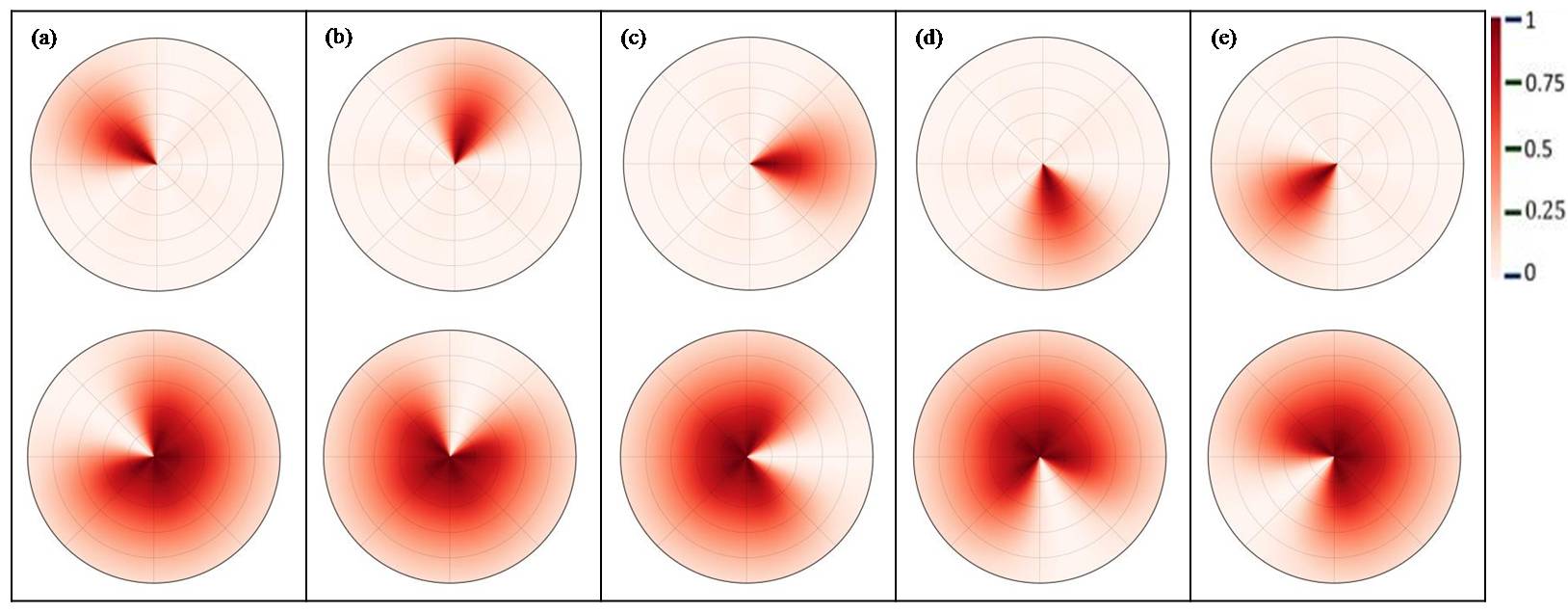}
    \caption{(a-e) Transverse plane intensity profiles of the output of a horizontally polarized Gaussian light beam, through each of the five d-plates of Fig.~\ref{Anglestates_Param}. The top and bottom rows correspond to horizontal and vertical components of the output vector beam.}
    \label{Intensity_Angle_states}
\end{figure}

Here, we give an illustration of the angle states generated from a horizontally polarized scalar beam through each of the plates depicted in Fig.~\ref{Anglestates_Param}.
\begin{figure}[H]
    \centering
    \includegraphics[width=1.0\linewidth]{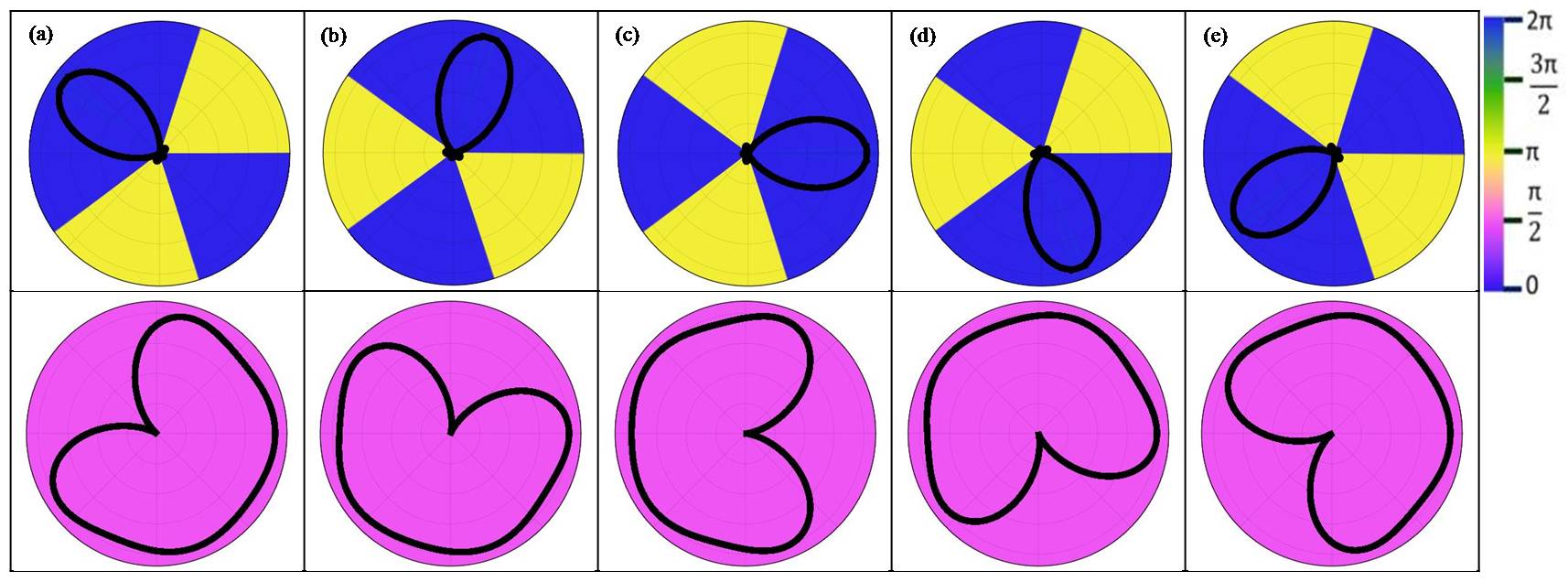}
    \caption{(a-e) The five angle states $\ket{2;j}$, for $j=-2,\dots,2$ respectively. The top and bottom rows represent the horizontal and vertical components of the vector beams.}
    \label{Anglestates_generation}
\end{figure}
Since the azimuthal functions corresponding to these angle states are all real, the phase in the horizontal component is only $0$ or $\pi$. The phase in the vertical component is uniform and is equal to $\frac{\pi}{2}$ by construction. The intensity in the horizontal component is $20$\% in all five cases. The transverse plane intensity profile in the horizontal and vertical components of these vector beams, with Gaussian intensity profile, is depicted in Fig.~\ref{Intensity_Angle_states}.

\section{\label{Sec:Appendix_Quantummaps} Quantum maps}
\begin{figure}[H]
    \centering
    \includegraphics[width=1.0\linewidth]{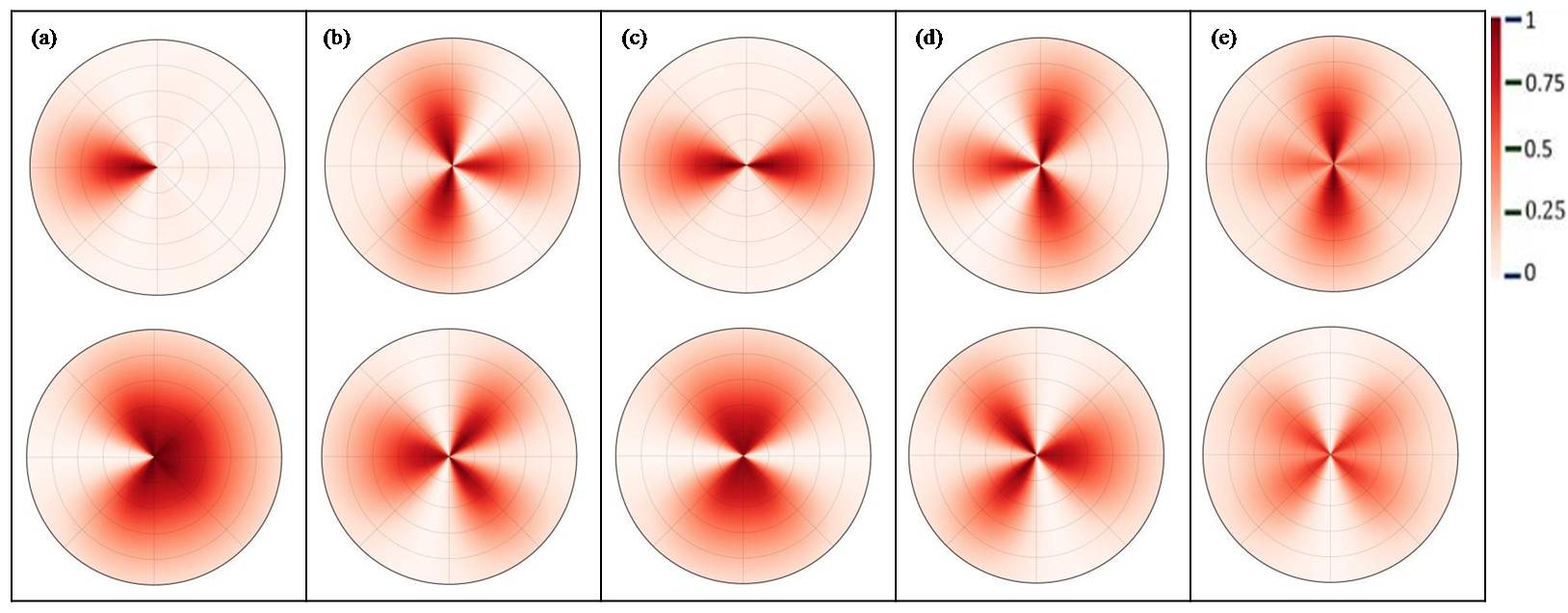}
    \caption{(a-e) Transverse plane intensity profiles of the output of a horizontally polarized Gaussian light beam, through each of the five d-plates of Fig.~\ref{OAMstates_Param}. The top and bottom rows correspond to horizontal and vertical components of the output vector beam.}
    \label{Intensity_OAM_states}
\end{figure}

We represent the OAM states defined in Eq.~\eqref{OAMStates_illus} generated from a scalar beam with horizontal polarization through each of the plates depicted in Fig.~\ref{OAMstates_Param}.
\begin{figure}[H]
    \centering
    \includegraphics[width=1.0\linewidth]{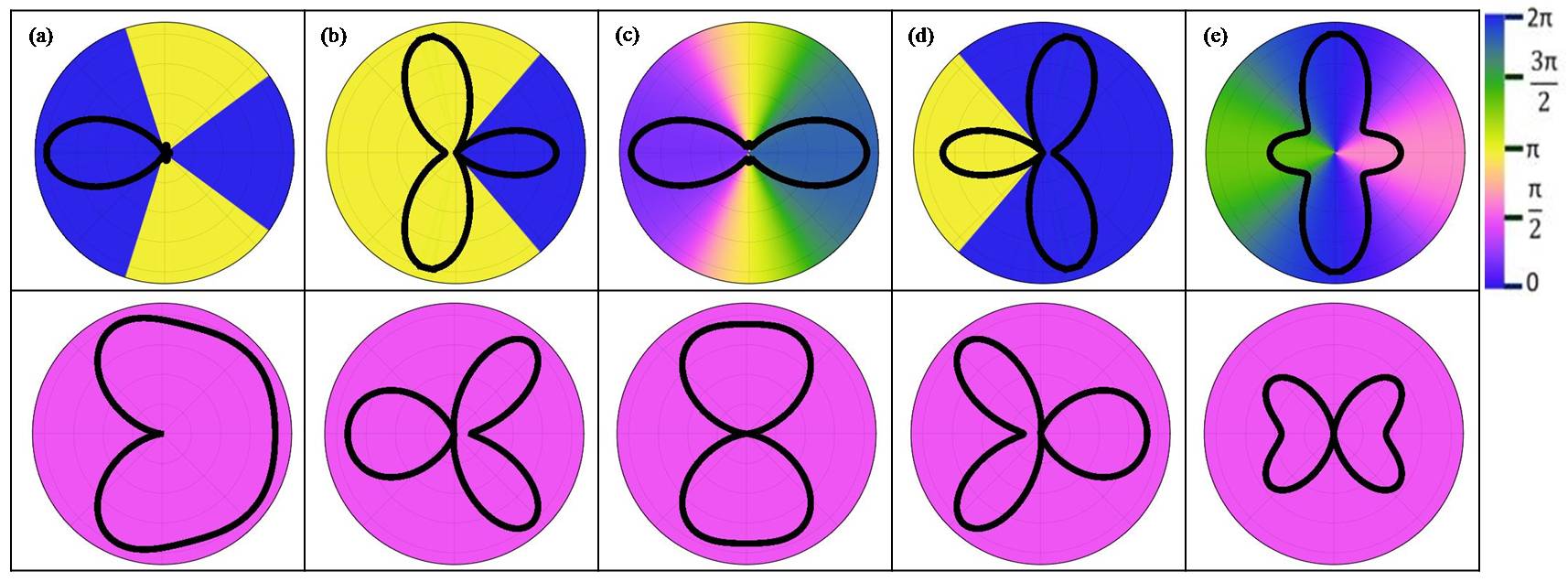}
    \caption{(a-e) The five OAM states $\ket{o_i}$, for $i=1,\dots,5$ of Eq.~\eqref{OAMStates_illus} respectively. The top and bottom rows represent the horizontal and vertical components of the emerging vector beams.}
    \label{OAMstates_generation}
\end{figure}
Since the functions corresponding to the OAM states $\ket{o_1}, \ket{o_2}$ and $\ket{o_4}$ are real, the phase in the horizontal component at every azimuth is either $0$ or $\pi$. The phase in the vertical component, on the other hand, is uniform for all the six states, and is equal to $\frac{\pi}{2}$ by construction. The transverse plane intensity profiles in the horizontal and vertical components of the vector beams represented in the above figure are depicted in Fig.~\ref{Intensity_OAM_states}.



\bibliography{VBT}

\end{document}